\documentclass[aps, showpacs, superscriptaddress, groupedaddress, twocolumn]{revtex4-1} 

\usepackage{amsmath}
\usepackage{amssymb}
\allowdisplaybreaks[2]

\usepackage{graphicx}
\usepackage{subfigure}

\usepackage{longtable} 

\usepackage{multirow}

\usepackage{dcolumn}
\usepackage{bm}
\usepackage{array}
\usepackage{xcolor} 

\usepackage{verbatim}

\begin{document}

\title{Long-range frustration in minimal vertex cover problem on random graphs}

\author{Yu-Tao Li$^1$}

\author{Chun-Yan Zhao$^2$}

\author{Jin-Hua Zhao$^3$}
\email{zhaojh@m.scnu.edu.cn}

\affiliation{$^1$College of Science,
University of Shanghai for Science and Technology,
Shanghai 200093, China}

\affiliation{$^2$School of Data Science and Engineering,
East China Normal University,
Shanghai 200062, China}

\affiliation{$^3$CoCoLab,
School of Data Science and Engineering,
South China Normal University,
Shanwei 516600, China}

\date{\today}


\begin{abstract}
A vertex cover on a graph is a set of vertices in which each edge of the graph is adjacent to at least one vertex in the set.
The minimal vertex cover (MVC) problem concerns finding vertex covers with the smallest cardinality, which is a typical computationally hard problem among combinatorial optimization on graphs.
Here, we follow the idea of the long-range frustration (LRF) in MVC configurations proposed in [\textsl{Physical Review Letters} \textbf{94} (2005) 217203].
We correct its analytical framework and further extend it from Erd\"{o}s-R\'enyi random graphs to general random graphs.
We formulate the framework of LRF into a percolation model, and analytically estimate the energy density of MVCs on uncorrelated random graphs only with their degree distributions.
We test our framework on some typical random graph models along with other methods, such as a hybrid algorithm of greedy leaf removal (GLR) procedure combined with survey propagation-guided decimation (SPD) algorithm and an analytical theory based on the GLR procedure which ignores LRF effect.
We show that, when there is a percolation of LRF effect, the above three predictions of energy density, say $x_{\rm LRF}$, $x_{\rm GLR + SPD}$, and $x_{\rm GLR}$, follow a scenario as $x_{\rm LRF} > x_{\rm GLR+SPD} > x_{\rm GLR}$ in most cases and $x_{\rm GLR+SPD} > x_{\rm LRF} > x_{\rm GLR}$ in the other cases, and $x_{\rm LRF}$ is much closer to $x_{\rm GLR+SPD}$ than $x_{\rm GLR}$ as $|x_{\rm LRF} - x_{\rm GLR+SPD} | < x_{\rm GLR+SPD} - x_{\rm GLR}$.
Our results show that LRF is a proper mechanism for the formation of complex energy landscape in the MVC problem and a theoretical framework of LRF helps to characterize its ground-state properties.


\end{abstract}

\maketitle


\section{Introduction}

A graph or a network \cite{Bollobas-2002, Boccaletti.etal-PhysRep-2006} is a simple language to describe the structure of interacted systems, which consists of vertices as their constituents and edges as interaction among constituents.
A typical combinatorial optimization problem defined on a graph \cite{Pardalos.Du.Graham-2013-2e} focuses on finding a set of vertices or edges with a minimal or maximal cardinality, in which certain constraints on the set are satisfied.
From the perspective of theoretical computer science, many combinatorial optimization problems are computationally hard \cite{Garey.Johnson-1979, Papadimitriou.Steiglitz-1998}, whose optimal solutions are buried in an exceedingly large solution space and it takes an unreasonable time (for example, to the order of an exponential function of a problem size) to find them in the worst case.
For these hard optimization problems, both developing fast algorithms for their solution configurations on problem instances and constructing an analytical theory for their ground-state properties prove to be difficult tasks.
A combinatorial optimization problem can be easily mapped to a statistical mechanics problem on a graph with discrete vertex states.
Statistical physics-based methods, especially the spin-glass theory, provide analytical and algorithmic tools, such as the replica trick and the cavity method \cite{Mezard.Parisi.Virasoro-1987, Nishimori-2001, Mezard.Montanari-2009, Zhou-2015}.
These approaches clarify the relation between the computational behavior of algorithms and the structural properties of solution space of underlying problems, and also develop fast message-passing algorithms on given graph instances.

An undirected graph $G = \{V, E\}$ has a vertex set $V$ with $|V| = N$ and an edge set $E$ with $|E| = M$. For any vertex $i \in V$, its degree $k_i$ is the size of the set of its nearest neighbors $\partial i$.
The mean degree of $G$ is $c \equiv 2M /N$.
The degree distribution $P(k)$ is defined as the probability of a randomly chosen vertex having a degree $k \geqslant 0$.
Another degree distribution important in analytical theories in the context of graphs is the excess degree distribution $Q(k)$. Following a randomly chosen edge $(i, j)$ between vertices $i$ and $j$, from $i$ to $j$, $Q(k)$ is defined as the probability that $j$ has a degree $k$. By definition, we have $Q(k) = k P(k) / c$.
As a classical model for random graphs, the Erd\"{o}s-R\'{e}nyi (ER) random graphs
\cite{Erdos.Renyi-PublMath-1959, Erdos.Renyi-Hungary-1960}
with a mean degree $c$ show a Poissonian degree distribution as
\begin{equation}
\label{eq:Poisson}
P(k) = {\rm e}^{- c} \frac{c^k}{k!},
\end{equation}
with $k \geqslant 0$.

A vertex cover on a graph $G = \{V, E\}$ is a set of vertices as $S$, such that each edge in the graph, say $(i, j) \in E$ between vertices $i$ and $j$, has at least one end-node in the set, say $i \in S$, or $j \in S$, or both $i, j \in S$.
A binary state for a vertex $i \in V$ can be defined as $s_i \in \{1, 0\}$, in which $s_i = 1$ denotes $i$ as being covered (in a vertex cover) and $s_i = 0$ as being uncovered (not in a vertex cover).
A microscopic configuration of vertex cover is $\vec{s} = \{s_i\}$ for all $i \in V$.
The constraint for $\vec{s}$ as a proper vertex cover can be stated as: for any edge $(i, j) \in E$, we have $(s_i, s_j)  = (1, 0)$, or $(0, 1)$, or $(1, 1)$.
The energy density or the fraction $x$ of a proper vertex cover configuration $\vec{s}$ is
\begin{align}
x = \frac {1}{N} \sum _{i \in V} s_i.
\end{align}
The minimal vertex cover (MVC) problem is to find vertex covers $S$ with the smallest cardinality, equivalently, those configurations with the lowest $x$.

The MVC problem can be further formulated as a statistical mechanics problem.
We first introduce $\beta$ as the inverse temperature.
The partition function of the MVC problem on $G = \{V, E\}$ is
\begin{align}
Z(\beta) = \sum _{\vec{s}} \prod _{i \in V} {\rm e}^{- \beta s_i} \prod _{(i, j) \in E} [1 - (1 - s_i) (1 - s_j)].
\end{align}
In the above equation, the first product is the Boltzmann factor of a covering configuration $\vec{s}$, and the second product selects these vertex cover configurations which satisfy the topological constraint.
$Z(\beta)$ simply sums all the proper vertex configurations and reweighs them with Boltzmann factors.
In the zero-temperature limit ($\beta \to + \infty$), only those $\vec{s}$ with the lowest energy contribute to the partition function.

Below we list some results, especially analytical ones, for the MVC problem in previous literature.
For more comprehensive reviews on statistical physics approaches to the MVC problem, interested readers can refer to \cite{Hartmann.Weigt-JPhyA-2003, Zhao.Zhou-CPB-2014}.

In the mathematical literature, an upper bound of the minimal energy density $x_0$ on a general graph $G = \{V, E\}$ based on vertex degrees is established in \cite{Harant-DiscreteMath-1998} as
\begin{align}
x_0 \leqslant 1 - \frac {1}{|V|} \frac {\left( \sum _{i \in V} \frac {1}{k_i + 1} \right)^2}
{\sum _{i \in V}\frac {1}{k_i + 1} - \sum _{(i, j)\in E} \frac {(k_i - k_j)^2}{(k_i + 1)(k_j + 1)}}.
\end{align}
On ER random graphs, upper and lower bounds for $x_0$ are derived in \cite{Gazmuri-Network-1984} as $x_l < x_0 < 1 - \ln c/c$, while $x_l$ is the root of
\begin{align}
x \ln x + (1 - x) \ln (1 - x) + \frac {c}{2}(1 - x)^2 = 0.
\end{align}
%
In \cite{Frieze-DiscreteMath-1990}, an asymptotic behavior of $x_0$ on ER random graphs with large $c$ is derived as
\begin{align}
x_0 = 1 - \frac {2}{c} \left( \ln c - \ln \ln c + 1 - \ln 2 \right) + o\left(\frac {1}{c} \right).
\end{align}
%

As a typical method inspired by statistical physics, the spin glass theory is also adopted for the MVC problem.
In \cite{Weigt.Hartmann-PRL-2000, Weigt.Hartmann-PRE-2001}, the replica trick is adopted and its energy density on ER random graphs under the assumption of replica symmetry (RS) is analytically calculated as
\begin{align}
\label{eq:x0-rs}
x_0 = 1 - \frac {W(c)}{c} - \frac {W^2{(c)}}{2c},
\end{align}
while the Lambert-W-function $W(c)$ is defined as $W(c) \exp W(c) = c$.
This prediction of $x_0$ is exact when $c \leqslant {\rm e} = 2.71828 \cdots$, yet it underestimates true ground-state energy density when $c > {\rm e}$.
When a RS solution fails, the picture of replica symmetry breaking (RSB) can be introduced for a more refined illustration of hierarchical structure of solution space.
In \cite{Weigt.Zhou-PRE-2006}, the survey propagation algorithm is developed  to characterize the properties of ground-state solutions at the first-step RSB level.
Yet whether the first-step RSB solution is enough or more steps of RSB are needed for the MVC problem on general random graphs is still an open question \cite{Zhang.Zeng.Zhou-PRE-2009, Barbier.etal-JPhysC-2013}.

In this paper, we focus on analytical frameworks for the MVC problem, which explicitly connect its ground-state properties with the structural parameters of underlying graphs.
These frameworks provide an explainable description to its ground-state properties, and act as a complimentary approach to those methods with a more algorithmic origin.
Yet constructing such frameworks is generally much more difficult than developing algorithms.
A frequently adopted approach is to explore the geometrical properties of solution configurations of optimization problems.
Below we briefly introduce two analytical frameworks: the first one can be considered as a baseline method for the analytical prediction of energy density of the MVC problem, and the second one is the framework we will correct and extend in this paper.

The first framework is based on the greedy leaf removal (GLR) procedure.
On an undirected graph, any vertex with a degree one is a leaf, and its only nearest neighbor is correspondingly a root.
The GLR procedure is the iterative removal of any root with all its adjacent edges, and leaves the residual subgraph as a core.
This procedure is originally adopted as a local algorithm to reduce problem size for the maximum matching (MM) problem \cite{Karp.Sipser-IEEFoCS-1981, Aronson.Frieze.Pittel-RandStrucAlgo-1998}, which finds matchings (a set of edges without shared vertices) with the maximal cardinality.
It is easy to verify that the roots constitute a part of a solution of MVC, and those edges connecting roots and their leaves belong to a part of a solution of MM.
The analytical theory of cores is developed on ER random graphs \cite{Bauer.Golinelli-EPJB-2001} and further on general random graphs \cite{Liu.Csoka.Zhou.Posfai-PRL-2012}.
With the analytical theory of both core and roots from the GLR procedure, the energy densities of the MVC and MM problems on general random graphs are further estimated \cite{Zhao.Zhou-JSTAT-2019}.
As it is shown for the MVC problem in \cite{Zhao.Zhou-JSTAT-2019}, when a core is absent, this theory gives a correct calculation of energy density. When there is a core, the trivial fixed point of the core percolation theory leads to an underestimation of energy density.
In \cite{Zhao.Zhou-JSTAT-2019}, equation (\ref{eq:x0-rs}) can be easily reproduced on ER random graphs with a simple geometrical interpretation.
A similar framework is also applied on the MM problem on the undirected bipartite representation of directed graphs \cite{Zhao.Zhou-PRE-2019}.
Generalized versions of the GLR procedure and their percolation analysis can also be found in other combinatorial optimization problems,
such as $k$-XORSAT problem \cite{Mezard.RicciTersenghi.Zecchina-JStatPhys-2003, Cocco.etal-PRL-2003},
Boolean networks \cite{Correale.etal-PRL-2006},
maximum independent set problem \cite{Lucibello.RicciTersenghi-IntJStatMech-2014},
minimum dominating set problem \cite{Zhao.Zhou-JStatPhys-2015, Zhao.Zhou-LNCS-2015},
covering problems on hypergraphs \cite{Coutinho.etal-PRL-2020},
and $z$-matching problem on bipartite graphs \cite{Zhao-JSTAT-2023}.
 
The second framework is based on the theory of long-range frustration (LRF) \cite{Zhou-PRL-2005, Zhou-PRL-2012}. The LRF effect among MVC configurations is based on an intuition that some combinations of states of distant vertex pairs are forbidden due to long paths between them.
Vertices whose states fluctuate among MVC configurations are further classified into two types, depending on how their state fixing triggers an extensive or a local state fixing of their neighboring vertices.
This framework provides a refined quantitative picture on how vertices in different coarse-grained states contribute to the energy density of the MVC problem.
Analytical result \cite{Zhou-PRL-2012} shows that the LRF theory on ER random graphs achieves estimations very close to those from survey propagation-guided decimation (SPD) algorithm, which is basically at the first-step RSB level.
The LRF framework is further applied on the satisfiability problems \cite{Zhou-NJP-2005, Zhou.Ma.Zhou-JSTAT-2007}.

In this paper, we follow the concept and the framework of LRF in \cite{Zhou-2015, Zhou-PRL-2005, Zhou-PRL-2012} and focus on the analytical theory of the energy density of the MVC problem on general random graphs.
Our contributions here are in three parts:
(1) we remove inconsistency and clarify some derivation steps in the LRF framework of \cite{Zhou-PRL-2005, Zhou-PRL-2012} which can lead to a significant deviation in the prediction of energy densities on random graphs with non-Poissonian degree distributions;
(2) we extend the corrected LRF framework on the MVC problem from the specific case of ER random graphs \cite{Zhou-PRL-2005, Zhou-PRL-2012} to sparse random graphs with arbitrary degree distributions;
(3) we test our LRF theory of the MVC problem on some random graph models, and it achieves predictions on energy densities close to the SPD algorithm, and proves to be better than an analytical theory based on the GLR procedure \cite{Zhao.Zhou-JSTAT-2019} which basically ignores LRF effect.

Here is the layout of the paper.
In section \ref{sec:model}, we present the concept of LRF for the MVC problem.
In section \ref{sec:theory}, we lay down our analytical framework of LRF on sparse random graphs.
In section \ref{sec:result}, we test our theory on some random graph models, and compare the results with three other algorithms.
In section \ref{sec:conclusion}, we conclude the paper with some discussion.

\section{Model}
\label{sec:model}

In this paper, we basically follow the concepts in \cite{Zhou-PRL-2005, Zhou-PRL-2012}, such as the classification of vertices based on MVC configurations, the notation of LRF, the distinction of type-I and type-II unfrozen vertices, and the definition of order parameter of LRF.
As shown in sections \ref{sec:theory} and \ref{sec:result}, we will make a substantial correction to the theoretical framework of LRF in \cite{Zhou-PRL-2005, Zhou-PRL-2012}. 
For the self-consistency of the paper, we simply lay down an intuitive picture and core ideas of LRF from \cite{Zhou-PRL-2005, Zhou-PRL-2012}.

\begin{figure}[htbp]
\begin{center}
 \includegraphics[width = 0.95 \linewidth]{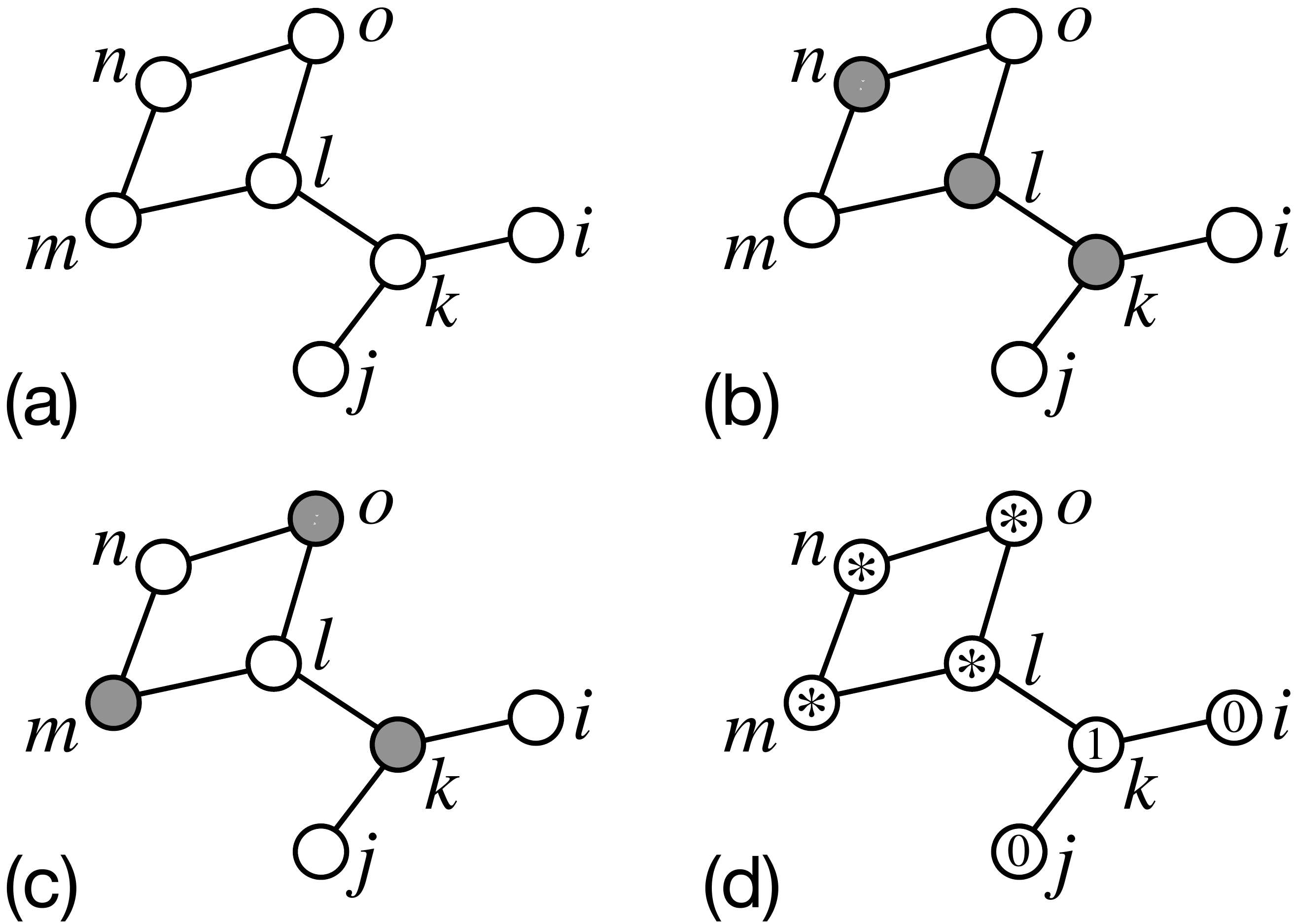}
\end{center}
\caption{
 \label{fig:model}
A diagram of the MVC configurations and the vertex categories.
(a) shows a small graph with $7$ vertices and $7$ edges.
(b) and (c) show two MVC configurations, in which covered vertices are denoted as shaded circles, and uncovered vertices are in empty circles.
(d) shows the categories of vertices based on the two MVC configurations, in which a vertex with $0$ inside is frozen as being uncovered, a vertex with $\ast$ inside is in an unfrozen state, and a vertex with $1$ inside is frozen as being covered.}
\end{figure}

As we mentioned, a MVC configuration on a graph $G = \{V, E\}$ can be denoted as $\vec{s} = \{s_i\}$ with $i \in V$ and $s_i \in \{0, 1\}$. 
For all the MVC configurations on $G$, there are simply three possibilities for the state of any vertex $i \in V$:
$s_i = 0$ for all configurations;
$s_i = 0$ for some (not all) configurations, while $s_i = 1$ for the other configurations; 
and $s_i = 1$ for all configurations.
We define a coarse-grained state $C = \{0, \ast, 1\}$ for the above three possibilities, respectively.
Correspondingly, we can classify all vertices into three categories: those frozen as being uncovered, those with an unfrozen state, and those frozen as being covered.
See figure \ref{fig:model} for an example.

On a large graph, we denote the relative sizes of vertices with coarse-grained states $\{0, \ast, 1\}$ as $(R_0, R_{\ast}, R_1)$, respectively. We simply have
\begin{equation}
R_0 + R_{\ast} + R_1 = 1.
\end{equation}
We can see that, vertices with the three coarse-grained states contribute to the energy of MVCs by $0$, a number $\in (0, 1)$, and $1$, respectively.
A quantitative description of $(R_0, R_{\ast}, R_1)$ helps to characterize energy density and other ground-state properties of the MVC problem.

The basic intuition of the LRF effect in the MVC problem is to further categorize those unfrozen vertices.
In figure \ref{fig:model}, the four vertices $\{l, m, n, o\}$ are all unfrozen vertices. For the non-neighboring vertex pairs $(l, n)$ and $(m, o)$, there is no combination for their states as $(1, 0)$ nor $(0, 1)$.
Such a case of forbidden state combinations can exist for two unfrozen vertices far apart on graphs, considering the typical distance $\propto \ln N$ between two vertices on a sparse graph with $N$ vertices.
To quantify the LRF effect, \cite{Zhou-PRL-2005, Zhou-PRL-2012} considers a simple question: when an unfrozen vertex $i$ is fixed with $s_i = 1$, how many unfrozen vertices must fix their states consequently.
Two simple scenarios are considered in \cite{Zhou-PRL-2005, Zhou-PRL-2012}:
a size of unfrozen vertices linear with $N$ fix their states correspondingly, thus the vertex $i$ is a type-I unfrozen vertex;
a finite size $\sim \mathcal{O}(1)$ of unfrozen vertices fix their states correspondingly, thus $i$ is a type-II unfrozen vertex.
On a large graph, the fraction $R$ of type-I unfrozen vertices among all unfrozen vertices is defined as the order parameter of the LRF effect \cite{Zhou-PRL-2005, Zhou-PRL-2012}.

In this paper, with a slightly different notation from the one in \cite{Zhou-PRL-2005, Zhou-PRL-2012}, we define the fraction of type-I unfrozen vertices among all vertices as $R_{\rm g}$.
In the following LRF theory for the MVC problem on random graphs, we develop an analytical framework to calculate $(R_0, R_{\ast}, R_{\rm g}, R_1)$ and finally the energy density $x$ of the MVC problem.

\section{Theory}
\label{sec:theory}

\subsection{Derivation of basic equations}

Our theory generally follows the one in \cite{Zhou-PRL-2005, Zhou-PRL-2012}.
As we will show, some inconsistencies of the theoretical framework in \cite{Zhou-PRL-2005, Zhou-PRL-2012} are removed to reach the current one in this paper.
We further extends the theory onto general random graphs with arbitrary degree distributions.
The framework here is basically the cavity method of spin glass theory, which is frequently applied on combinatorial optimization problems and satisfiability problems \cite{Mezard.Montanari-2009, Zhou-2015}, and also on percolation models on graphs \cite{Newman.Strogatz.Watts-PRE-2001, Li.etal-PhysRep-2021}.
Despite drastically different equations of the cavity method for specific problems on graphs, the cavity method has a probabilistic formalism. Some basic quantities of problems, such as the ground-state energy density, can be interpreted as marginal probabilities on a graph.
Under a given anzatz of RS or levels of RSB, these marginal probabilities can be derived with a set of cavity probabilities and those cavity probabilities form self-consistent equations with each other.

On a sparse random graph $G = \{V, E\}$, the relative sizes of vertices $(R_0, R_{\ast}, R_{\rm g}, R_1)$ can be understood as marginal probabilities when a randomly chosen vertex in $V$ is frozen as being uncovered, unfrozen, type-I unfrozen, and frozen as being covered, respectively.
These marginal probabilities can be calculated with pertinent cavity probabilities.
On a randomly chosen edge $(i, j) \in E$ and from vertex $i$ to vertex $j$, we define two cavity probabilities $(r_{0}, r_{\rm g})$: $r_{0}$ as the probability of $j$ frozen as being uncovered and $r_{\rm g}$ as the probability of $j$ being a type-I unfrozen vertex, both when $(i, j)$ is not considered.
For a randomly chosen vertex $i \in V$, its category depends on these categories of its nearest neighbors.
We consider here a cavity graph $G \backslash i$ when $i$ and its adjacent edges are all removed from $G$.
Under the Bethe-Peierls approximation \cite{Mezard.Montanari-2009} at the RS level in cavity method, the categories of $i$'s nearest neighbors in $G \backslash i$ are independent with each other due to long distances between them.
We then can establish equations connecting marginal probabilities $(R_0, R_{\ast}, R_{\rm g}, R_1)$ and cavity probabilities $(r_0, r_{\rm g})$.

In figure \ref{fig:equations}, we show a schematics to summarize the main notations and the flowchart in our theoretical framework. A detailed explanation is laid down in the following paragraphs.

\begin{widetext}


%
\begin{figure}[htbp]
\begin{center}
 \includegraphics[width = 0.99 \linewidth]{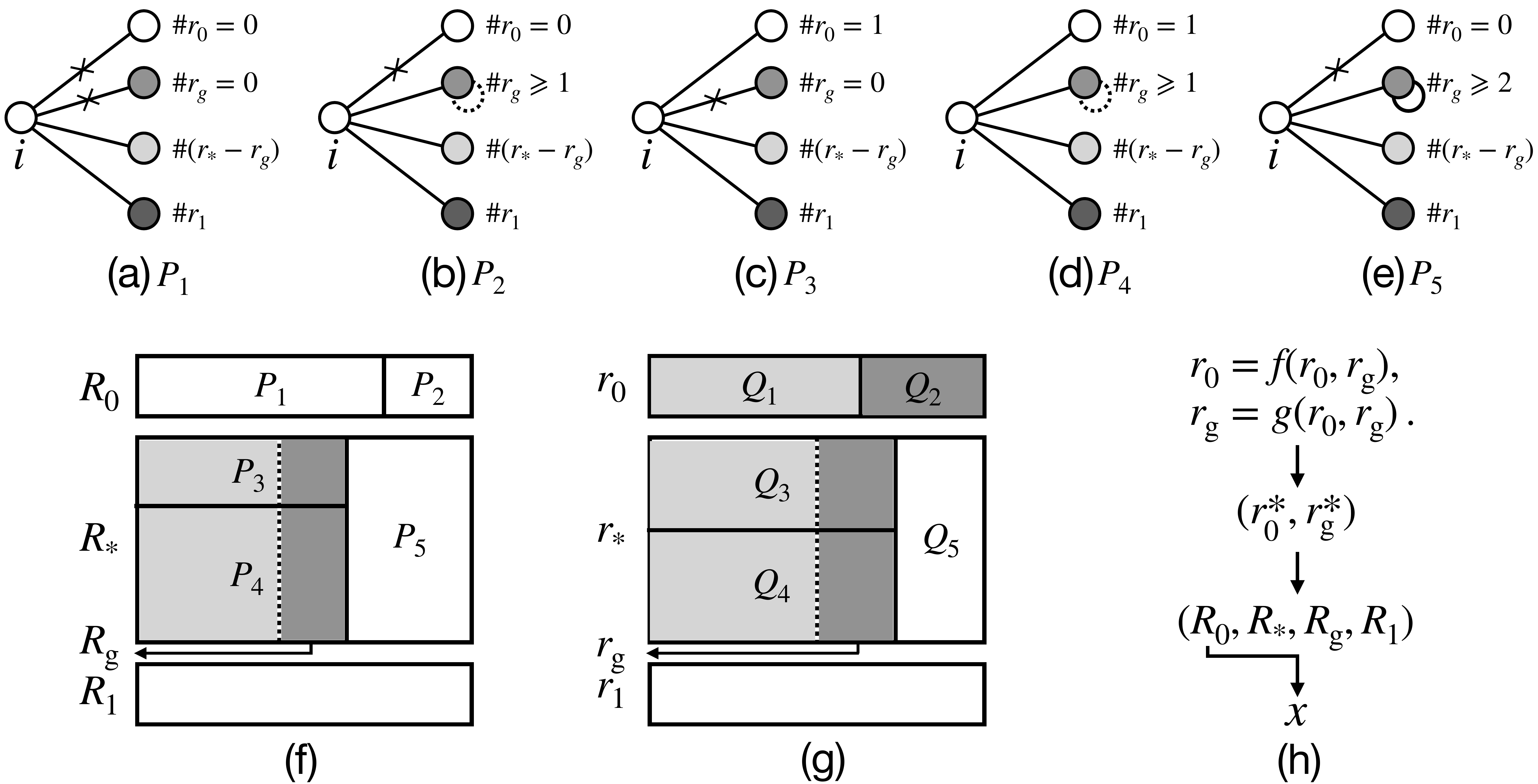}
\end{center}
\caption{
 \label{fig:equations}
A diagram of the main notations of probabilities and the flowchart of calculation in the LRF theory.
(a)--(e) shows a vertex $i$ and the five different cases of the states of its nearest neighbors.
All of the frozen neighbors in the uncovered state, the type-I unfrozen neighbors, the type-II unfrozen neighbors, and the frozen neighbors in the covered states are denoted collectively with an empty circle, a dark shaded circle, a light shaded circle, and a filled circle, respectively.
Notations of $\# r_0$, $\# r_{\rm g}$, $\# (r_{\ast} - r_{\rm g})$,  and $\# r_1$ respectively denote the size of the four types of neighboring vertices.
A cross on an edge reiterates the absence of neighbors of a certain type.
The five subfigures gives us a guide to the derivation of $P_1$ in equation (\ref{eq:P1}), $P_2$ in equation (\ref{eq:P2}), $P_3$ in equation (\ref{eq:P3}), $P_4$ in equation (\ref{eq:P4}), and $P_5$ in equation (\ref{eq:P5}), respectively.
In (b) and (d), a dashed circle touching a dark shaded circle denotes the situation when all the $\# r_{\rm g}$ type-I unfrozen neighbors are not frustrated and can be in the covered state at the same time.
In (e), a solid circle touching a dark shaded circle denotes the situation when at most $\# r_{\rm g} - 1$ type-I unfrozen neighbors are not frustrated and can be in the covered state at the same time.
(f) and (g) show the components of the marginal probabilities $(R_0, R_{\ast}, R_{1})$ and the cavity probabilities $(r_0, r_{\ast}, r_{1})$, respectively.
$R_{\rm g}$ and $r_{\rm g}$ correspond to the dark shaded sections in $R_{\ast}$ of (f) and in $r_{\ast}$ of (g), respectively.
The ratios of light and dark shaded areas in $R_{\ast}$ of (f) and in $r_{\ast}$ of (g) both equal to $Q_1/Q_2$ in (g), as shown in equations (\ref{eq:Rg}) and (\ref{eq:rg}), respectively.
(h) shows the flowchart from the self-consistent equations of $r_0$ and $r_{\rm g}$ to the calculation of the energy density $x$.
A detailed explanation is in section \ref{sec:theory}.}
\end{figure}
%

We first consider $R_0$.
For a randomly chosen vertex $i \in V$ to be frozen as being uncovered, we consider two cases for the nearest neighbors $\partial i$ in $G \backslash i$ as Case I and Case II.
In Case I, among $\partial i$ there is no vertex frozen as being uncovered, nor type-I unfrozen vertex.
The probability of this case is
\begin{align}
\label{eq:P1}
P_1
& = \sum _{k = 0}^{+ \infty} P(k) (1 - r_0 - r_{\rm g})^{k}.
\end{align}
%

In Case II, among $\partial i$ there is no vertex frozen as being uncovered, yet with at least one type-I unfrozen vertex. Yet there is a possibility that all these type-I unfrozen vertices are not frustrated in $G \backslash i$, thus they can be in the covered state in certain MVC configurations.
For any two type-I unfrozen vertices in $\partial i$ in $G \backslash i$, we assume an equal chance for them to be frustrated and be not frustrated, neglecting their local structural properties.
Then the probability for Case II is
\begin{align}
\label{eq:P2}
P_2
& = \sum _{k = 1}^{+ \infty}
P(k) \sum _{s = 1}^{k} {k \choose s} r _{\rm g}^s (1 - r_0 - r_{\rm g})^{k - s} \frac{1}{2^{s - 1}} \\
\label{eq:P2-simplified}
& = 2 \sum _{k = 0}^{+ \infty} P(k) \left[
\left( 1 - r_0 - \frac {r_{\rm g}}{2} \right)^{k} - (1 - r_0 - r_{\rm g})^{k} \right].
\end{align}
From equation (\ref{eq:P2}) to equation (\ref{eq:P2-simplified}), we simply substitute the summation sign $\sum _{s = 1}^{k}$ with $\sum _{s = 0}^{k}$, and further rearrange the equation into concise summations on degree distributions. The form of summations in equation (\ref{eq:P2-simplified}) is suitable for short-hand notations as we will show later.

In both Cases I and II, when a vertex $i$ is added into $G \backslash i$, $i$ becomes a vertex frozen as being uncovered in $G$.
Thus we have $R_0$ as
\begin{align}
R_0 = P_1 + P_2.
\end{align}
With equations (\ref{eq:P1}) and (\ref{eq:P2-simplified}), we have $R_0$ equivalently as
\begin{align}
\label{eq:R0-simplified}
R_0
= 2 \sum _{k = 0}^{+\infty} P(k) \left( 1 - r_0 - \frac {r_{\rm g}}{2} \right)^{k}
- \sum _{k = 0}^{+ \infty} P(k) (1 - r_0 - r_{\rm g})^{k}.
\end{align}
%

Beware that, even though in both Case I and Case II $i$ becomes a vertex frozen as being uncovered, their influence on $i$'s neighbors are significantly different.
In Case I, since there is no type-I unfrozen vertex among $\partial i$ in $G \backslash i$, after $i$ is assigned as being uncovered in $G$, there is only a limited number of unfrozen neighbors which are assigned with a certain state in the subsequent state fixing.
Yet in Case II, a state fixing from $i$ can propagate until a macroscopic fraction of unfrozen vertices is assigned with certain states.
We can see that a drastic change happens in the configurations of the MVC problem.

Then we consider $R_{\ast}$.
For a randomly chosen vertex $i \in G$, we consider three cases, Case III, Case IV, and Case V, for $\partial i$ in $G \backslash i$.
In Case III, among $\partial i$ there is only one vertex frozen as being uncovered and no type-I unfrozen vertex. We have the probability term as
\begin{align}
\label{eq:P3}
P_3 = \sum _{k = 1}^{+ \infty} P(k) k r_0 (1 - r_0 - r_{\rm g})^{k - 1}.
\end{align}

In Case IV, among $\partial i$ there is only one vertex frozen as being uncovered and at least one type-I unfrozen vertex. Yet there is a possibility that these type-I unfrozen neighbors are not frustrated in $G \backslash i$, thus they can be all in the covered state, like the situation in Case II.
We have the probability term as
\begin{align}
\label{eq:P4}
P_4
& = \sum _{k  = 2}^{+ \infty} P(k) k r_0 \sum _{s = 1}^{k - 1}
{k - 1 \choose s} r_{\rm g}^{s} (1 - r_0 - r_{\rm g})^{k - 1 - s} \frac {1}{2^{s - 1}} \\
\label{eq:P4-simplified}
& = 2 r_0 \sum _{k = 1}^{+ \infty} P(k) k
\left[ \left( 1 - r_0 - \frac {r_{\rm g}}{2} \right)^{k - 1}
- (1 - r_0 - r_{\rm g})^{k - 1} \right].
\end{align}
From equation (\ref{eq:P4}) to equation (\ref{eq:P4-simplified}), we substitute the second summation sign $\sum _{s = 1}^{k - 1}$ with $\sum _{s = 0}^{k - 1}$ and rearrange the equation into simple summations forms.

In Case V, among $\partial i$ there is no vertex frozen as being uncovered and at least two type-I unfrozen vertices. Suppose here we have $s (\geqslant 2)$ type-I unfrozen vertices in $\partial i$ in $G \backslash i$. There is a possibility that a frustration shows between one vertex and all the other $s - 1$ vertices while these $s - 1$ vertices are not frustrated, thus at most $s - 1$ neighbors can be in the covered state.
Following the logic in Cases II and IV, we have the probability term as
\begin{align}
\label{eq:P5}
P_5
& = \sum _{k = 2}^{+ \infty} P(k)
\sum _{s = 2}^{k} {k \choose s} r_{\rm g}^{s} (1 - r_0 - r_{\rm g})^{k - s}
\left[ \delta _{s,2} \frac {1}{2} + (1 - \delta _{s,2}) \frac {s}{2^{s - 1}} \right] \\
\label{eq:P5-simplified}
& = cr_{\rm g}  \sum _{k = 1}^{+\infty} Q(k)
\left[ \left( 1 - r_0 - \frac {r_{\rm g}}{2} \right)^{k - 1}
- (1 - r_0 - r_{\rm g})^{k - 1} \right]
 - \frac {c r_{\rm g}^2}{4} \sum _{k = 2}^{+\infty} Q(k) (k - 1) (1 - r_0 - r_{\rm g})^{k - 2}.
\end{align}
In equation (\ref{eq:P5}), the Kronecker delta $\delta _{s, t}$ is defined as $1$ only when $s = t$ and $0$ in other cases.
From equation (\ref{eq:P5}) to equation (\ref{eq:P5-simplified}), we leave the details of simplification in appendix A.

In all the Cases III, IV, and V, a vertex $i$ becomes an unfrozen vertex in $G$. Thus we have
\begin{align}
R_{\ast} = P_3 + P_4 + P_5.
\end{align}
With equations (\ref{eq:P3}), (\ref{eq:P4-simplified}), and (\ref{eq:P5-simplified}), we correspondingly have
\begin{align}
\label{eq:Ra-simplified}
R_{\ast}
& = (2 c r_0 + c r_{\rm g}) \sum _{k = 1}^{+\infty} Q(k)
\left( 1 - r_0 - \frac {r_{\rm g}}{2} \right)^{k - 1} 
- (c r_0 + c r_{\rm g}) \sum _{k = 1}^{+\infty} Q(k)
(1 - r_0 - r_{\rm g})^{k - 1} \nonumber \\
& - \frac {c r_{\rm g}^2}{4} \sum _{k = 2}^{+\infty} Q(k) (k - 1) (1 - r_0 - r_{\rm g})^{k - 2}.
\end{align}
Details of above equation are also left in appendix A.

Then we consider $R_{\rm g}$.
We can see that only in Cases III and IV, the vertex $i$ can be type-I unfrozen in $G$, yet an extra constraint should be satisfied.
Here we denote a vertex $j \in \partial i$ as the only vertex which is frozen as being uncovered in $G \backslash i$.
After the addition of $i$ into $G \backslash i$, $j$ correspondingly becomes an unfrozen vertex in $G$.
If $i$ is assigned as $s_i = 1$ in $G$, we have a state fixing as $s_j = 0$ to achieve a low energy.
For the vertex $j$ per se, there are Cases I and II for the nearest neighbors $\partial j \backslash i$, which lead to $j$ as frozen as being uncovered before the addition of $i$.
If the nearest neighbors $\partial j \backslash i$ is in Case II, a propagation of state fixing happens and a macroscopic fraction of unfrozen vertices are assigned with certain states.
For a randomly chosen edge $(i, j) \in E$ between vertices $i$ and $j$, we define $Q_1$ and $Q_2$ as the probability of Case I and Case II, respectively, for the states of $\partial j \backslash i$. We consider them as the cavity counterpart of marginal probabilities $P_1$ and $P_2$, respectively.
Following equations (\ref{eq:P1}) - (\ref{eq:P2-simplified}) for the derivation of $P_1$ and $P_2$, we lay down expressions for $Q_1$ and $Q_2$ as
\begin{align}
\label{eq:Q1}
Q_1
& = \sum _{k = 1}^{+ \infty} Q(k) (1 - r_0 - r_{\rm g})^{k - 1}, \\
\label{eq:Q2-orginal}
Q_2
& = \sum _{k = 2}^{+ \infty} Q(k)
\sum _{s = 1}^{k - 1} {k - 1 \choose s} r_{\rm g}^{s} (1 - r_0 - r_{\rm g})^{k - 1 - s} \frac {1}{2^{s - 1}} \\
\label{eq:Q2-simplified}
& = 2 \sum _{k = 1}^{+ \infty} Q(k)
\left[ \left( 1 - r_0 - \frac {r_{\rm g}}{2} \right)^{k - 1}
- (1 - r_0 - r_{\rm g})^{k - 1}
\right].
\end{align}
We thus have
\begin{align}
r_0 = Q_1 + Q_2.
\end{align}
Combining equations (\ref{eq:Q1}) and (\ref{eq:Q2-simplified}), we have
\begin{align}
\label{eq:r0-simplified}
r_0 = 2 \sum _{k = 1}^{+ \infty} Q(k) \left(1 - r_0 - \frac {r_{\rm g}}{2} \right)^{k - 1}
- \sum _{k = 1}^{+ \infty} Q(k) \left(1 - r_0 - r_{\rm g} \right)^{k - 1}.
\end{align}
%

For $R_{\rm g}$, we have
\begin{align}
\label{eq:Rg}
R_{\rm g}
= (P_3 + P_4) \frac {Q_2}{Q_1 + Q_2}
= (P_3 + P_4) \left(1 - \frac {Q_1}{r_0} \right).
\end{align}
After inserting equations (\ref{eq:P3}), (\ref{eq:P4-simplified}), and (\ref{eq:Q1}), we finally have
\begin{align}
\label{eq:Rg-simplified}
R_{\rm g}
& = \left[ 2 \sum _{k = 1}^{+ \infty} P(k) k \left(1 - r_0 - \frac {r_{\rm g}}{2} \right)^{k - 1}
- \sum _{k = 1}^{+ \infty} P(k) k \left(1 - r_0 - r_{\rm g} \right)^{k - 1}
\right] 
 \left[ r_0 - \sum _{k = 1}^{+ \infty} Q(k) (1 - r_0 - r_{\rm g})^{k - 1}
\right].
\end{align}
%

We finally consider $r_{\rm g}$.
For a randomly chosen edge $(i, j) \in E$ between vertices $i$ and $j$, we consider Case III and Case IV for the states of $\partial j \backslash i$. Their probabilities are defined as $Q_3$ and $Q_4$, respectively, which can be viewed as the cavity counterpart of marginal probabilities $P_3$ and $P_4$, respectively.
Following equations (\ref{eq:P3}) - (\ref{eq:P4-simplified}) for the derivation of $P_3$ and $P_4$, we lay down expressions for $Q_3$ and $Q_4$ as
\begin{align}
\label{eq:Q3}
Q_3
& = \sum _{k = 2}^{+ \infty} Q(k) (k - 1) r_0 (1 - r_0 - r_{\rm g})^{k - 2}, \\
\label{eq:Q4}
Q_4
& = \sum _{k  = 3}^{+ \infty} Q(k) (k - 1) r_0 \sum _{s = 1}^{k - 2}
{k - 2 \choose s} r_{\rm g}^{s} (1 - r_0 - r_{\rm g})^{k - 2 - s} \frac {1}{2^{s - 1}} \\
\label{eq:Q4-simplified}
& = 2 r_0 \sum _{k = 2}^{+ \infty} Q(k) (k - 1)
\left[ \left( 1 - r_0 - \frac {r_{\rm g}}{2} \right)^{k - 2}
- (1 - r_0 - r_{\rm g})^{k - 2} \right].
\end{align}
%

Following equation (\ref{eq:Rg}) for the derivation of $R_{\rm g}$, we have the expression for $r_{\rm g}$ as
\begin{align}
\label{eq:rg}
r_{\rm g}
= (Q_3 + Q_4) \frac {Q_2}{Q_1 + Q_2}
= (Q_3 + Q_4) \left(1 - \frac {Q_1}{r_0} \right).
\end{align}
Combining equations (\ref{eq:Q1}), (\ref{eq:Q3}), and (\ref{eq:Q4-simplified}), we have
\begin{align}
\label{eq:rg-simplified}
r_{\rm g}
& = \left[
2 \sum _{k = 2}^{+ \infty} Q(k) (k - 1) \left(1 - r_0 - \frac {r_{\rm g}}{2} \right)^{k - 2}
- \sum _{k = 2}^{+ \infty} Q(k) (k - 1) \left(1 - r_0 - r_{\rm g} \right)^{k - 2}
\right] 
\left[
r_0 - \sum _{k = 1}^{+ \infty} Q(k) (1 - r_0 - r_{\rm g})^{k - 1}
\right].
\end{align}

\end{widetext}

Equations (\ref{eq:R0-simplified}), (\ref{eq:Ra-simplified}), (\ref{eq:Rg-simplified}), (\ref{eq:r0-simplified}), and (\ref{eq:rg-simplified}) consist of the basis of our analytical theory.
For a graph ensemble or instance with $P(k)$, with equations (\ref{eq:r0-simplified}) and (\ref{eq:rg-simplified}) we first calculate stable fixed $(r_0, r_{\rm g})$ , then with equations (\ref{eq:R0-simplified}), (\ref{eq:Ra-simplified}), and (\ref{eq:Rg-simplified}) we calculate corresponding $(R_{0}, R_{\ast}, R_{\rm g})$, respectively.

Here is a numerical procedure to solve the stable fixed points of equations (\ref{eq:r0-simplified}) and (\ref{eq:rg-simplified}).
We define the right-hand side of equations (\ref{eq:r0-simplified}) and (\ref{eq:rg-simplified}) as $f(r_{0}, r_{\rm g})$
and $g(r_{0}, r_{\rm g})$, respectively. We have
\begin{align}
\label{eq:r0-f}
r_{0} & = f(r_{0}, r_{\rm g}), \\
\label{eq:rg-g}
r_{\rm g} & = g(r_{0}, r_{\rm g}).
\end{align}
We then adopt a numerical method to calculate stable fixed points of $r_{0}$ and $r_{\rm g}$.
For any given $r_{\rm g} \in [0, 1]$, we calculate the stable fixed point of $r_{0}$ with equation (\ref{eq:r0-f}) as $r_{0}^{\ast}$.
Then we can calculate $g(r_{0}^{\ast}, r_{\rm g})$.
Finally, we can check whether $r_{\rm g} = g(r_{0}^{\ast}, r_{\rm g})$ satisfies.
In such a way, we can calculate stable fixed $r_{\rm g}$ for equation (\ref{eq:rg-g}) as $r_{\rm g}^{\ast}$. Then we again calculate corresponding $r_{0}^{\ast}$ with given $r_{\rm g}^{\ast}$ with equation (\ref{eq:r0-f}).
Finally, we have the pair of fixed points as $(r_{0}^{\ast}, r_{\rm g}^{\ast})$.

With the solution of $(R_0, R_{\ast}, R_{\rm g})$, we can further estimate ground-state properties of the MVC problem.
We focus on its energy density $x$.
Consider here a vertex $i$ is added into a cavity graph $G \backslash i$.
If $i$ becomes a vertex frozen as being uncovered in $G$, the addition of $i$ contributes by no energy to MVC on $G$.
If $i$ becomes an unfrozen vertex in $G$, its sole uncovered nearest neighbor (say, $j$) also becomes an unfrozen vertex in $G$. Then either $i$ or $j$ is covered in the MVC on $G$.  Thus the addition of $i$ contributes by one to the energy of MVC on $G$.
If $i$ becomes a frozen vertex as being covered in $G$, there is no state change in $G \backslash i$, and the addition of $i$ also contributes by one to the energy of MVC on $G$.
Summing the above three cases, $x$ can be estimated as \cite{Zhou-PRL-2012}
\begin{align}
\label{eq:energy-density}
x = \frac {1}{c} \int _{0}^{c} (1 - R_0) {\rm d}c'.
\end{align}
To calculate $x$ for a given $c$, we first discretize $c$ as $c = N_c \Delta c$ into $N_c (\gg 1)$ steps with a step length $\Delta c$.
Thus we have a long sequence of mean degrees $c'$ as $c' = n \Delta c$ with $1 \leqslant n \leqslant N_c$.
We should mention that, for each $c'$,  the corresponding $R_0$ is calculated from our framework.
We calculate $R_0$ for each $c'$, and add up $1 - R_0$ as the energy contribution from $c' - \Delta c$ to $c'$.
Beware that, the above procedure takes an incremental approach to calculate the energy density on large graphs with a mean degree $c$, which effectively calculates all the energy densities of MVC with $c' < c$.
It is easy to see that, with a larger $N_c$, we can have a more accurate $x$.

Here we discuss the time complexity of our analytical theory to estimate energy density $x$ for a graph ensemble or instance with a degree distribution $P(k)$ and a corresponding mean degree $c$.
To calculate the stable fixed $(r_{0}^{\ast}, r_{\rm g}^{\ast})$, we adopt a simple bisection method to find all fixed points of $(r_{0}, r_{\rm g})$ in a greedy way. In the search for fixed points, we discretize $[0, 1]$ (the range of $r_0$ and $r_{\rm g}$) into $N_0$ equal intervals for $r_0$ and $N_{\rm g}$ equal intervals for $r_{\rm g}$. Thus the time complexity to find the stable fixed $(r_{0}^{\ast}, r_{\rm g}^{\ast})$ is $\mathcal{O} (N_0 N_{\rm g})$.
Considering the discretization scheme in estimating energy density in equation (\ref{eq:energy-density}), the final time complexity of our framework to calculate the energy density at mean degree $c$, along with those at each mean degree $c' (\leqslant c)$ in the discretization step, is $\mathcal {O}(N_c N_0 N_{\rm g})$.
In section \ref{sec:result}, we set $N_c = 10^3$ for ER random graphs and $10^4$ for the other graph models, and $N_0 = N_{\rm g} = 10^3$ uniformly.

\subsection{Comparison with the theory based on the GLR procedure}

We then compare our framework with the theory based on the GLR procedure for the MVC problem in \cite{Zhao.Zhou-JSTAT-2019}.
We first make a very crude assumption that there is no LRF among unfrozen vertices, and we see where our analytical framework will lead us.
In this context, we have $r_{\rm g} = R_{\rm g} = 0$.
Our framework reduces to
\begin{align}
\label{eq:r0-no-LRF}
r_0
&= \sum _{k = 1}^{+\infty} Q(k) (1 - r_0)^{k - 1}, \\
R_0
&= \sum _{k = 0}^{+\infty} P(k) (1 - r_0)^{k}, \\
R_{\ast}
&= cr_0 \sum _{k = 1}^{+\infty} Q(k) (1 - r_0)^{k - 1} \nonumber \\
& = cr_0^2, \\
R_1
&= 1 - R_0 - R_{\ast} \nonumber \\
&= 1 -  \sum _{k = 0}^{+\infty} P(k) (1 - r_0)^{k} - cr_0^2.
\end{align}
We further assume that each type-II vertex contributes by exactly one half to the energy of the MVC problem, neglecting details of local graph structure.
As there are only type-II vertices in unfrozen vertices here, their total contribution to the energy density is simply $R_{\ast} / 2$. We have the energy density as
\begin{align}
\label{eq:x-no-LRF}
x = \frac{R_{\ast}}{2} + R_1 = 1 -  \sum _{k = 0}^{+\infty} P(k) (1 - r_0)^{k} - \frac {1}{2} cr_0^2.
\end{align}
Given a degree distribution $P(k)$, with equation (\ref{eq:r0-no-LRF}) we first calculate stable fixed $r_0$, and then with equation (\ref{eq:x-no-LRF}) we can estimate $x$.
On the other hand, in \cite{Zhao.Zhou-JSTAT-2019} to estimate the energy density of the MVC problem, the branch of trivial fixed points of $\alpha$ and $\beta$ with $1 - \alpha - \beta = 0$ is chosen no matter there is a core or not on a graph.
It is easy to find some correspondence of notations and equations between our work and \cite{Zhao.Zhou-JSTAT-2019}:
the cavity probability $r_0$ to the cavity probability $\alpha$, equation (\ref{eq:r0-no-LRF}) to equations (1) and (2) when $1 - \alpha - \beta = 0$, and equation (\ref{eq:x-no-LRF}) to equation (5).
Taken the above messages together, the theory based on the GLR procedure simply corresponds to the LRF theory with no LRF effect. When a macroscopic LRF effect sets in and $R > 0$, this theory deviates from a proper prediction and underestimates the energy density of MVCs.

\subsection{Simplified equations of our theory}

Our basic equations can be further reformulated in a more compact form.
We first define two short-hand summations on degree distributions as
\begin{align}
\label{eq:sumPk}
P^{(s)} (x)
& = \sum _{k = s}^{+ \infty} P(k) {k \choose s} x^{k - s},\\
\label{eq:sumQk}
Q^{(s)}(x)
& = \sum _{k = s + 1}^{+ \infty} Q(k) {k - 1 \choose s} x^{k - 1 - s},
\end{align}
in which $x \in [0, 1]$ is a real variable and $s \geqslant 0$ is an integer.
Equations (\ref{eq:R0-simplified}), (\ref{eq:Ra-simplified}), (\ref{eq:Rg-simplified}),
(\ref{eq:r0-simplified}), and (\ref{eq:rg-simplified}) can be rewritten as
\begin{align}
R_{0}
& = 2 P^{(0)} \left(1 - r_{0} - \frac {r_{\rm g}}{2} \right)
- P^{(0)} \left(1 - r_{0} - r_{\rm g}\right), \\
R_{\ast}
& = (2 c r_0 + cr_{\rm g}) Q^{(0)} \left(1 - r_0 - \frac {r_{\rm g}}{2} \right) \nonumber \\
& - (c r_0 + cr_{\rm g}) Q^{(0)}(1 - r_0 - r_{\rm g})
- \frac {c r_{\rm g}^2}{4} Q^{(1)}(1 - r_0 - r_{\rm g}), \\
R_{\rm g}
& = \left[ 2 P^{(1)} \left(1 - r_{0} - \frac {r_{\rm g}}{2} \right)
- P^{(1)} \left(1 - r_{0} - r_{\rm g} \right) \right] \nonumber \\
& \times
\left[ r_{0} - Q^{(0)} \left( 1 - r_{0} - r_{\rm g} \right) \right],\\
r_{0}
& = 2 Q^{(0)} \left(1 - r_{0} - \frac {r_{\rm g}}{2} \right)
- Q^{(0)} \left(1 - r_{0} - r_{\rm g}\right),\\
r_{\rm g}
& = \left[ 2 Q^{(1)} \left(1 - r_{0} - \frac {r_{\rm g}}{2} \right)
- Q^{(1)} \left(1 - r_{0} - r_{\rm g} \right) \right] \nonumber \\
& \times \left[ r_{0} - Q^{(0)} \left( 1 - r_{0} - r_{\rm g} \right) \right].
\end{align}
%

\section{Result}
\label{sec:result}

\subsection{Methods for comparison}

To further ascertain the correctness of energy density prediction from our framework, we also calculate energy densities of the MVC problem with other three methods.

The first method is the theory based on the GLR procedure \cite{Zhao.Zhou-JSTAT-2019}.
We leave a simple explanation of this analytical method in appendix B.
Predictions from the GLR-based theory will show us how the neglection of LRF effect results in an underestimation of energy density of the MVC problem.

The second method is the belief propagation-guided decimation (BPD) algorithm \cite{Zhao.Zhou-CPB-2014} combined with the GLR procedure, which outputs approximate MVC configurations on graph instances.
We simply name it as the GLR+BPD algorithm for short.
The belief propagation (BP) algorithm works at the RS level, assuming that all the solutions are organized in a single cluster (a macroscopic state) which has no inner structure. The inverse temperature $\beta$ is the reweighting parameter in the BP algorithm.
The basic procedure of this hybrid algorithm is as follows:
(1) on a graph instance, we first apply the GLR procedure to cover roots as local optimal steps until there is a core;
(2) we iterate cavity messages on all edges of the residual core, until a convergence of messages or a maximal number of message updating;
(3) we adopt a BPD step on the core to cover a fraction of vertices with the largest marginal probability to be in the covered state;
(4) the above three steps are iteratively carried out, until all the edges of the initial graph are covered.
Basic parameters of message updating and vertex decimation in the algorithm are as follows:
the maximal iteration number $N_{\rm iter}$ in a single step of message updating,
the criterion $\varepsilon$ of message convergence for the maximal difference between messages between two consecutive updating steps, 
and the size of covered vertices $N_{1}$ in a single decimation step of BPD on a core with a vertex size $N_{\rm core}$.
In our result here, we set $N_{\rm iter} = 200$, $\varepsilon = 10^{-8}$, $N_{1} = \max \{N_{\rm core}/ N_{\rm d}, N_{\rm min}\}$ with $N_{\rm d} = 200$ and $N_{\rm min} = 1$.

The third method is the SPD algorithm \cite{Weigt.Zhou-PRE-2006} combined with the GLR procedure, which also outputs approximate MVC configurations on graph instances.
We name it as the GLR+SPD algorithm for short.
The survey propagation (SP) algorithm \cite{Mezard.Parisi.Zecchina-Science-2002} works at the first-step RSB level, assuming that the solutions are organized in a large number of well separated clusters, in each of which there is no inner structure.
In each macroscopic state, we set $\beta \to +\infty$, and reweigh different macroscopic states by their energy densities with a parameter $y$.
The procedure of the GLR+SPD algorithm is much like the one in the GLR+BPD algorithm, in which we iterate messages based on equations in SP algorithm rather than those based on equations in BP algorithm.
In our result here, we adopt the same parameters of $N_{\rm iter}$, $\varepsilon$, $N_{1}$, $N_{\rm d}$, and $N_{\rm min} $ with the GLR+BPD algorithm.

\subsection{Results on random graph models}

Then we test our analytical framework on some representative random graph models.

We first consider the ER random graphs
\cite{Erdos.Renyi-PublMath-1959, Erdos.Renyi-Hungary-1960}.
For Poissonian degree distributions, the summations in equations (\ref{eq:sumPk}) and (\ref{eq:sumQk}) reduce to
\begin{align}
\label{eq:er-ps-qs}
P^{(s)}(x) = Q^{(s)}(x) = {\rm e}^{- c (1 - x)} \frac {c^s}{s!},
\end{align}
for any $x \in [0, 1]$ and $s \geqslant 0$.
We have simplified formulae for $(R_{0}, R_{\ast}, R_{\rm g}, r_{0}, r_{\rm g})$ as
\begin{align}
\label{eq:R0-er}
R_{0}
& = 2 {\rm e}^{- c(r_0 + r_{\rm g} / 2)} - {\rm e}^{- c (r_{0} + r_{\rm g})}, \\
\label{eq:Ra-er}
R_{\ast}
& = (2 c r_0 + c r_{\rm g}) {\rm e}^{- c \left( r_0 + r_{\rm g} / 2 \right)} \nonumber \\
& - \left[ cr_0 + cr_{\rm g} + \frac {(cr_{\rm g})^2}{4} \right] {\rm e}^{- c (r_0 + r_{\rm g})}, \\
\label{eq:Rg-er}
R_{\rm g}
& = \left[ 2 c {\rm e}^{- c(r_0 + r_{\rm g} / 2)} - c {\rm e}^{- c (r_{0} + r_{\rm g})} \right]
\left[ r_0 - {\rm e}^{- c (r_{0} + r_{\rm g})} \right], \\
\label{eq:r0-er}
r_{0}
& = 2 {\rm e}^{- c (r_0 + r_{\rm g} / 2)} - {\rm e}^{- c (r_{0} + r_{\rm g})}, \\
\label{eq:rg-er}
r_{\rm g}
& = \left[ 2 c {\rm e}^{- c (r_0 + r_{\rm g} / 2)} - c {\rm e}^{- c (r_{0} + r_{\rm g})} \right]
\left[ r_0 - {\rm e}^{- c (r_{0} + r_{\rm g})} \right].
\end{align}
We can easily find that
\begin{align}
\label{eq:R0r0-Rgrg}
R_{0} = r_0, R_{\rm g} = r_{\rm g}.
\end{align}
This equivalence is simply a natural result of equation (\ref{eq:er-ps-qs}) specifically on ER random graphs, which is not a universal property for general random graphs.

We then discuss connections and difference between our framework and Zhou's in \cite{Zhou-PRL-2005, Zhou-PRL-2012}.
In \cite{Zhou-PRL-2005}, equation (2) shows
\begin{align}
\label{eq:Zhou-2005-r0}
q_{+} = 2 {\rm e}^{- c q_{+} - c q_{0}R / 2} - {\rm e}^{- c q_{+} - c q_{0} R},
\end{align}
and equation (3) shows
\begin{align}
\label{eq:Zhou-2005-ra}
q_0
& = (2 c q_{+} + c q_0 R) {\rm e}^{- c q_{+} - c q_0 R / 2} \nonumber \\
& - \left[ c q_{+} + c q_0 R + \frac{(c q_0 R)^2}{4} \right] {\rm e}^{- c q_{+} - c q_0 R}.
\end{align}
The first equation in \cite{Zhou-PRL-2012} shows
\begin{align}
\label{eq:Zhou-2012-rg}
R = \frac {c q_{+}^2}{q_{0}} \left( 1 - \frac {1}{q_{+}} {\rm e}^{- c q_{+} - c q_{0} R} \right).
\end{align}
Considering a correspondence between probabilities as $\{q_{+}, q_{0}R\} \leftrightarrow \{r_0, r_{\rm g}\}$, we can easily find that the right-hand side of equations (\ref{eq:Zhou-2005-r0}) , (\ref{eq:Zhou-2005-ra}), and (\ref{eq:Zhou-2012-rg}) is equivalent to equations (\ref{eq:R0-er}), (\ref{eq:Ra-er}), and (\ref{eq:Rg-er}), respectively.
This means that our framework naturally reproduces those main results in \cite{Zhou-PRL-2005, Zhou-PRL-2012}.
Yet there is a fundamental difference between these two frameworks.
Considering the forms of degree distributions, the right-hand side of both equation (2) in \cite{Zhou-PRL-2005} and the first equation in \cite{Zhou-PRL-2012} are meant for marginal probabilities (say $R_0$ and $R_{\rm g}$ in our notations here), not for cavity probabilities (say $r_0$ and $r_{\rm g}$).
That is to say, Zhou's theory establishes its framework only with marginal probabilities. Our theory reproduces Zhou's one only when equation (\ref{eq:R0r0-Rgrg}) holds on ER random graphs, and Zhou's theory can suffer from severe deviation on predictions on graphs with non-Poissonian degree distributions.
In summary, our framework takes the LRF effect as intrinsically a percolation model and follows a formal construction process for its analytical theory with both marginal and cavity probabilities, thus we can root out the intrinsic inconsistencies in Zhou's theory.

%
\begin{figure*}[htbp]
\begin{center}
 \includegraphics[width = 0.95 \linewidth]{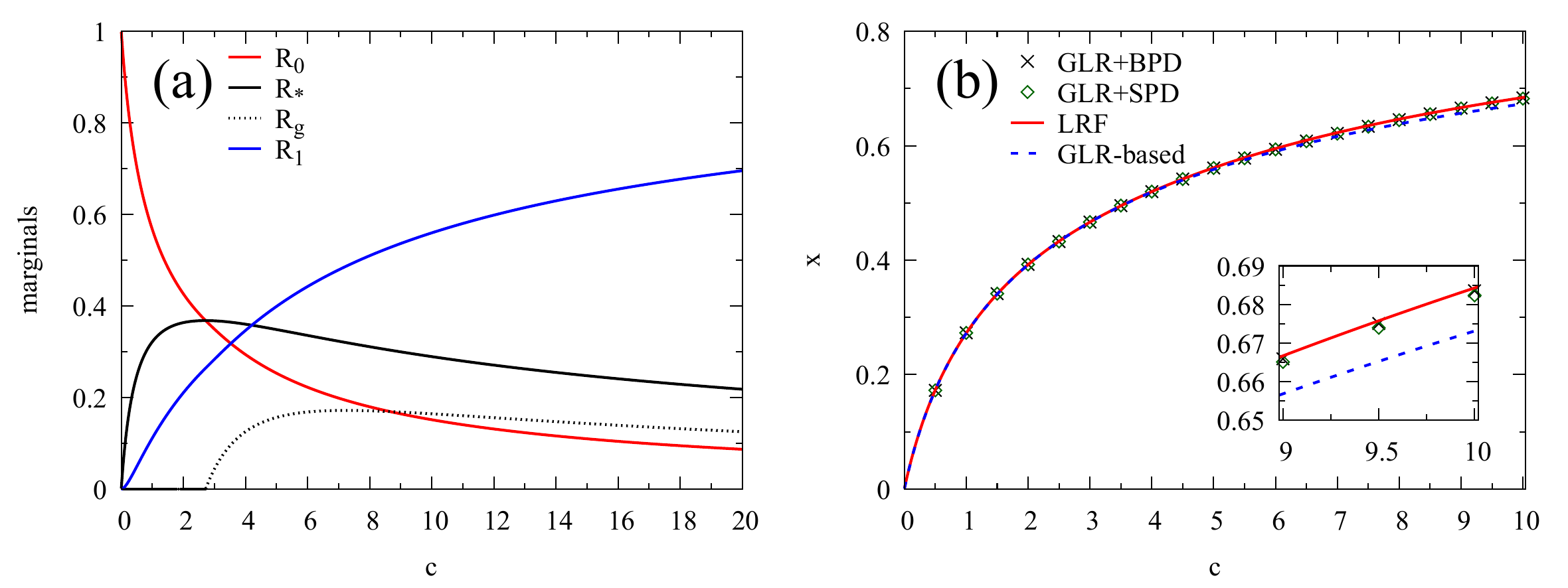} 
\end{center}
\caption{
 \label{fig:er}
Marginals of the LRF theory and energy densities of the MVC problem on ER random graphs.
(a) shows marginal probabilities $(R_0, R_{\ast}, R_{\rm g}, R_1)$ from the LRF theory on infinitely large ER random graphs.
(b) shows energy densities of the MVC problem on ER random graphs.
Results are from four methods:
the GLR+BPD algorithm on graph instances with a vertex size $N = 10^5$ with $\beta = 10$,
the GLR+SPD algorithm on graph instances with a vertex size $N = 10^5$ with $y = 3.1$,
the framework of the LRF theory on infinitely large graphs with $\Delta c = 0.001$,
and the theory based on the GLR procedure (GLR-based) on infinitely large graphs.}
\end{figure*}
%

In figure \ref{fig:er}(a), we can see that with an increasing $c$, $R_0$ decreases and $R_1$ increases monotonously. $R_{\ast}$ gradually increases from $0$, reaches a maximum at $c \approx {\rm e}$, and then decreases. $R_{\rm g}$ follows a similar pattern with $R_{\ast}$, yet a nontrivial $R_{\rm g} (> 0)$ emerges continuously at $c = c^{\ast} = {\rm e}$, while $c^{\ast}$ is the critical mean degree.
The behavior of $R_{\rm g}$ can be explained by the birth of a nontrivial stable $r_{\rm g}$ from equations (\ref{eq:r0-er}) and (\ref{eq:rg-er}).
In a simple schematic explanation, we can plot $y = g(r_0, r_{\rm g})$ and $y = r_{\rm g}$ in the same coordinates, and all the fixed points can be read from the intersection between them.
When $c \leqslant c^{\ast} = {\rm e}$, there is only one fixed point as $r_{\rm g}^{\ast} = 0$.
When $c > c^{\ast} = {\rm e}$, a nontrival stable point $r_{\rm g}^{\ast} (> 0)$ emerges continuously.
In a general case, the critical point $c^{\ast}$ can be determined from the set of equations as
\begin{align}
g(r_0, r_{\rm g}) |_{c = c^{\ast}} & = 0, \\
\frac {\partial g(r_0, r_{\rm g})}{\partial r_{\rm g}} |_{c = c^{\ast}}& = 1.
\end{align}
%

In figure \ref{fig:er}(b), we show energy densities of the MVC problem from four methods.
When $c \leqslant {\rm e}$, there is no percolation of type-I unfrozen vertices, equivalently no LRF effect among unfrozen vertices, and all the four methods achieve energy densities with an indistinguishable difference.
When $c > {\rm e}$, there is a percolation of LRF effect among unfrozen vertices, and we have the following three observations.
The first one is that, the GLR+SPD algorithm achieves consistently lower energy densities than the GLR+BPD algorithm.
Thus we take the GLR+SPD results as a convenient reference of true ground-state energy density of the MVC problem.
The second one is that, the LRF predictions are higher than the GLR+SPD results, and the GLR-based predictions are lower than the GLR+SPD results.
An intuitive explanation is that the effect of LRF forbids some covering configurations which have lower energies yet violate the structural constraint for a proper vertex cover. Ignoring the LRF effect simply leads to an underestimation of true ground-state energy density of the MVC problem.
The third one is that, with the GLR+SPD results as a reference, the LRF predictions are much more closer to them than the GLR-based predictions.
The three complementary observations confirm that the idea of LRF captures a proper mechanism leading to the highly complicated energy landscape of the MVC problem, and our theoretical framework on LRF delivers a reasonable prediction of its ground-state energy densities.


%
\begin{figure*}[htbp]
\begin{center}
 \includegraphics[width = 0.90 \linewidth]{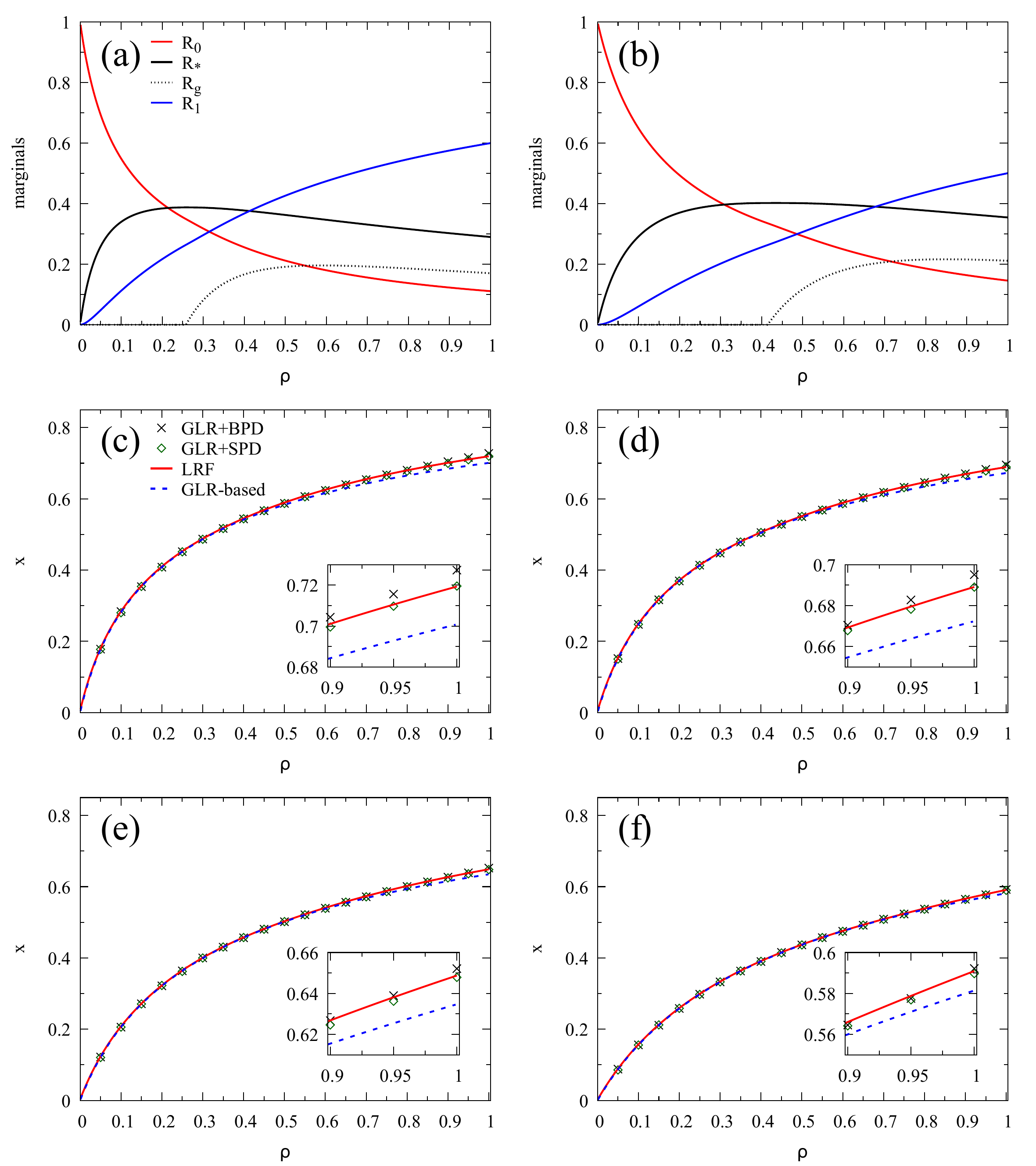} 
\end{center}
\caption{
 \label{fig:rr}
Marginals of the LRF theory and energy densities of the MVC problem on diluted RR graphs.
(a) and (b) show marginal probabilities $(R_0, R_{\ast}, R_{\rm g}, R_1\}$ from the LRF theory on infinitely large diluted RR graphs in the case of $K = 10$ and $K = 6$ , respectively.
(c)--(f) show energy densities of the MVC problem on diluted RR graphs from four methods in the case of $K = \{10, 8, 6, 4\}$, respectively.
Each subgraph in (c)--(f) generally follows the format and the algorithm parameters in figure \ref{fig:er}(b).
For the GLR+SPD algorithm, we set $y = 3$.
For the LRF theory, we set $\Delta \rho = 0.001$.}
\end{figure*}
%

We then test our framework on diluted regular random (RR) graphs.
A RR graph has a uniform degree distribution as each vertex has a degree $K (\geqslant 2)$. In order to generate graph instances with a heterogeneous degree profile, we randomly dilute a RR graph, in which a fraction $1 - \rho \in [0, 1]$ of edges is randomly chosen and removed. The residual diluted RR graph shows a degree distribution $P(k)$ as
\begin{align}
P(k) = {K \choose k} \rho ^{k} (1 - \rho)^{K - k},
\end{align}
with $0 \leqslant k \leqslant K$.

In figure \ref{fig:rr}, the four marginal probabilities for diluted RR graphs follow the similar pattern as the case on ER random graphs. 
We find that when a percolation happens, the LRF predictions of energy densities are only slightly higher than the GLR+SPD results, and they are much closer to the GLR+SPD results than the GLR-based predictions.


%
\begin{figure*}[htbp]
\begin{center}
 \includegraphics[width = 0.90 \linewidth]{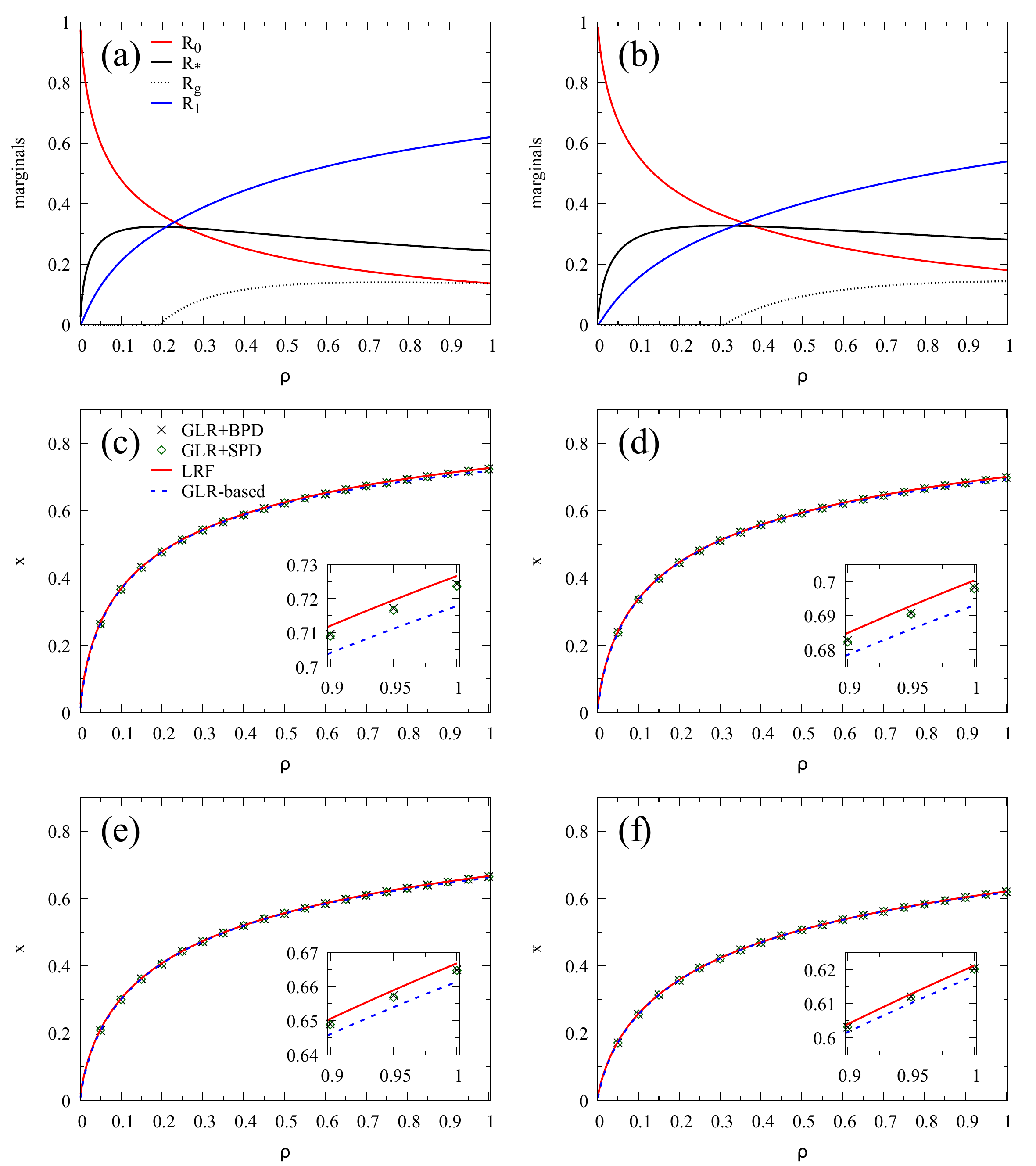} 
\end{center}
\caption{
 \label{fig:sf}
Marginals of the LRF theory and energy densities of the MVC problem on scale-free networks generated with configurational model.
We here focus on network instances generated with $N = 10^5$, $\gamma = 2.5$, and $k_{\rm max} = \sqrt {N}$.
(a) and (b) show marginal probabilities $(R_0, R_{\ast}, R_{\rm g}, R_1\}$ from the LRF theory on a single graph instance in the case of $k_{\rm min} = 12$ and $k_{\rm  min} = 8$, respectively.
(c)--(f) show energy densities of the MVC problem from four methods on diluted graph instances in the case of $k_{\rm min} = \{12, 10, 8, 6\}$, respectively.
Each subgraph in (c)--(f) generally follows the format and the algorithm parameters in figure \ref{fig:er}(b).
For the LRF and the GLR-based theoretical frameworks, the empirical degree distribution of diluted graph instances is the input.
For the LRF theory, we set $\Delta \rho = 0.001$.}
\end{figure*}
\begin{figure*}[htbp]
\begin{center}
 \includegraphics[width = 0.90 \linewidth]{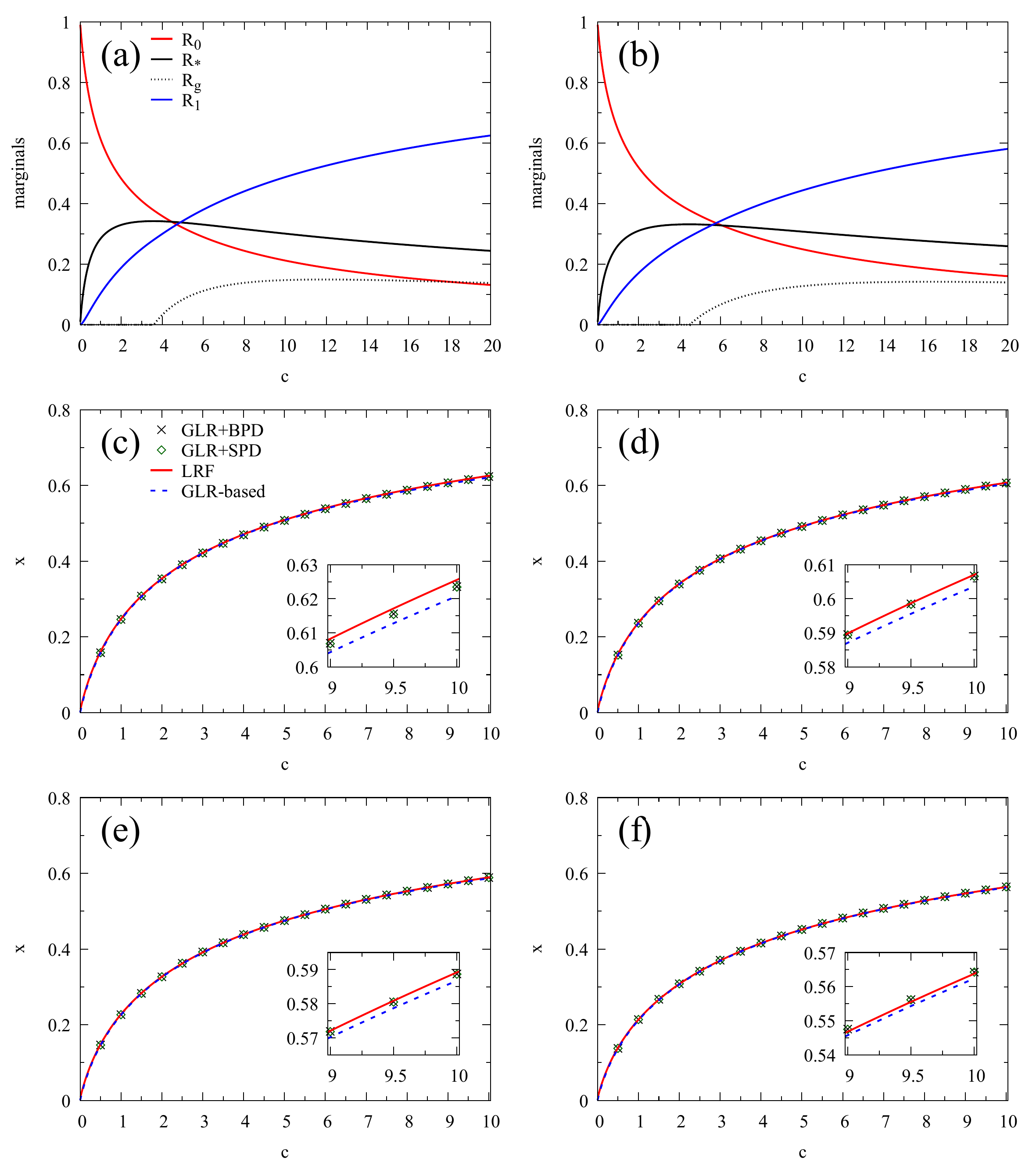} 
\end{center}
\caption{
 \label{fig:sm}
Marginals of the LRF theory and energy densities of the MVC problem on scale-free networks generated with static model.
(a) and (b) show marginal probabilities $(R_0, R_{\ast}, R_{\rm g}, R_1)$ from the LRF theory on inifitely large scale-free networks in the case of $\gamma = 3.5$ and $\gamma = 3$, respectively.
(c)--(f) show energy densities of the MVC problem from four methods on scale-free networks in the case of $\gamma = \{3.5, 3.2, 3, 2.8\}$, respectively.
Each subgraph in (c)--(f) generally follows the format and the algorithm parameters in figure \ref{fig:er}(b).
For the LRF theory, we set $\Delta c = 0.01$.}
\end{figure*}
%

We finally consider networks with scale-free property in degree distributions \cite{Barabasi.Albert-Science-1999}, which show a degree distribution $P(k) \propto k^{- \gamma}$ with $\gamma$ as a degree exponent.
This property exists abundantly in real-world networks due to their intricate evolution mechanisms, which lead to structural heterogeneity at many different levels.

Here we consider two models to generate scale-free network instances.
The first model is the configurational model \cite{Newman.Strogatz.Watts-PRE-2001}, which can generate graph instances with any given proper degree distribution.
A typical procedure of the configurational model follows as: for a degree distribution $P(k)$ and a vertex size $N$, we have a list of degrees $k$ and corresponding vertex size $NP(k)$; we generate a sequence of degrees with a size of $N$, in which a vertex with a degree $k$ has $k$ half-edges; two half-edges from two different vertices can be connected into a proper edge if there is no edge between them; after all half-edges are turned into proper edges, we finally generate a graph instance.
In the configurational model for scale-free networks, we define four parameters: a vertex size $N$, a degree exponent $\gamma$, a maximal degree $k_{\rm max}$, and a minimal degree $k_{\rm min}$.
To eliminate the degree-degree correlation in networks, we usually set $k_{\rm max} = \sqrt N$.
Like the case in diluted RR graphs, we consider the diluted version of scale-free network instances, in which a fraction $1 - \rho \in [0, 1]$ of edges is randomly chosen and removed.
For a scale-free network instance with an initial degree distribution $P^{i}(k)$ with $k_{\rm min} \leqslant k \leqslant k_{\rm max}$, after a dilution process with a fraction $\rho$, we have the degree distribution of the diluted graph as
\begin{align}
P(k) = \sum _{t = \max \{k_{\rm min}, k \}}^{k_{\rm max}} P^{i}(t) {t \choose k} \rho ^{k} (1 - \rho)^{t - k},
\end{align}
with $0 \leqslant k \leqslant k_{\rm max}$.

In figure \ref{fig:sf}, the four marginal probabilities of LRF theory on a scale-free network instance follow a pattern quite similar to those in the case of ER random graphs.
We find that when there is a percolation, the LRF predictions of energy densities are always higher than the GLR+SPD results.
We further notice that, with the GLR+SPD results as a reference, even the overestimation from LRF theory is comparable with the underestimation from the GLR-based theory, the LRF predictions are still closer to the GLR+SPD results.

We then consider scale-free networks generated with static model \cite{Goh.Kahng.Kim-PRL-2001, Catanzaro.PastorSatorras-EPJB-2005}.
The graph construction process in static model is a process of independent edge addition based on weights of vertices, which is much like the generation of ER random graphs.
Basic parameters of the model are a vertex size $N$, a degree exponent $\gamma$, and a mean degree $c$.
We first define an intermediary parameter $\xi  \equiv 1 / (\gamma - 1)$.
We initially construct a null graph with only $N$ vertices and no edge.
For each vertex $i$ with an index $i \in \{1, 2, \cdots, N\}$, a relative weight $w_i = i^{- \xi} / \sum _{i = 1}^{N} i^{- \xi}$ is assigned. In a single step of edge addition, two vertices, say $i$ and $j$, are selected with probabilities of their respective weights $w_i$ and $w_j$. If there is no connection between them, a proper edge between them can be established as $(i, j)$.
In such a way, a sequence of edges with a size $M = c N / 2$ is added into the null graph.
Such a large graph instance has a degree distribution as
\begin{eqnarray}
P(k) = \frac {1}{\xi}\frac {[c (1 - \xi)]^k}{ k!} {\rm E}_{- k + 1 + \frac{1}{\xi}} (c (1 - \xi)),
\end{eqnarray}
with $k \geqslant 0$.
The special function ${\rm E}_{a}(x)$ is a general exponential integral function defined as
${\rm E}_{a}(x) \equiv \int _{1}^{\infty} \mathrm{d}t e^{-xt} t^{- a}$
with $a, x > 0$. For large $k$, we have $P (k) \propto k^{- \gamma}$.

In figure \ref{fig:sm}, we find that the pattern of the four marginals is also quite similar to the case of ER random graphs.
We also find that when a percolation happens, the LRF predictions of energy densities are slightly higher than the GLR+SPD results except the case of $\gamma = 2.8 (< 3.0)$, and the GLR-based predictions are lower than the GLR+SPD results. Besides, the LRF predictions are always closer to the GLR+SPD results than the GLR-based predictions.

\section{Conclusion}
\label{sec:conclusion}

In this paper, we focus on an analytical method for the MVC problem, which predicts its energy density as an explicit function of the topological properties of graph ensembles.
This methodology is complementary to those powerful methods, such as SP algorithm, which mainly apply on graph instances or are in a formalism of population dynamics.

Our theoretical framework is based on the effect of LRF between unfrozen vertices in MVC configurations.
We correct the LRF theory in \cite{Zhou-PRL-2005, Zhou-PRL-2012} and extend it from ER random graphs onto general random graphs.
An analytical theory on LRF is finally developed to account for the relative size of vertices in the four coarse-grained states, thus leads to a theoretical prediction on the energy density of MVC only with degree distribution of graphs as inputs.
We test our analytical framework on some random graph models.
We show that, when a percolation of LRF effect happens on a graph, the performance of our framework is rather close to a hybrid algorithm combining the GLR procedure and the SPD algorithm, and is significantly better than a theory based on the GLR procedure.
That is to say, our analytical framework of LRF at the RS level can achieve predictions of energy densities very close to the message passing algorithms at the first-step RSB level.
Our framework shows that a refined picture of the structure of solution configurations helps us to develop a precise theory for ground-state properties of hard combinatorial optimizations.

We should mention that our binary classification of unfrozen vertices is only an approximation on the path to an exact theory, if possible, of the energy density of the MVC problem on general random graphs.
More detailed information on the structure of unfrozen vertices can further improve our prediction.
For example, we can incorporate the size distribution of state fixing from an unfrozen vertex into our current framework. Potential improvement on our theory will be carried out in a future work.

The concept of LRF is based on the long-range interaction among vertex states in the ground-state configurations of optimization problems, and the LRF effect is not limited to the context of the MVC problem.
\cite{Zhou-NJP-2005} introduces the LRF framework into $K$-satisfiability problems to discuss the percolation thresholds of the LRF effect with the SAT-UNSAT transition points.
\cite{Zhou.Ma.Zhou-JSTAT-2007} calculates the residual order parameter of LRF in the first-step RSB solutions for the MVC and the maximal $2$-satisfiability problems.
Both are examples of a new way to elucidate the phase transition behaviors in the solution spaces of combinatorial optimization problems from the perspective of the spin glass theory.
An integration of the LRF effect into other hard optimization problems can be considered in a following work.

\section*{Acknowledgements}

J-H Zhao thanks Prof. Hai-Jun Zhou (ITP-CAS) for discussions.
The authors thank the anonymous reviewers for helpful comments.
J-H Zhao is supported by Guangdong Basic and Applied Basic Research Foundation of China (Grant No. 2022A1515011765) and National Natural Science Foundation of China (Grant No. T2541021).
C-Y Zhao is supported by the National Key Research and Development Program of China (No. 2023YFC3304700), the Shanghai 2024 “Science and Technology Innovation Action Plan” Project (No. 24BC3201100), and the Program of Shanghai Academic/Technology Research Leader (No. 23XD1401100).

\begin{widetext}

\section*{Appendix A: Simplification of $P_5$ and $R_{\ast}$}

\setcounter{equation}{0}
\renewcommand{\theequation}{A.\arabic{equation}}

Here we list the details in the simplification of $P_5$ in equation (\ref{eq:P5-simplified})  and $R_{\ast}$ in equation (\ref{eq:Ra-simplified}).

For $P_5$, we have
\begin{align}
P_{5}
& = \sum _{k = 2}^{+\infty} P(k)
\left[ {k \choose 2} r_{\rm g}^2 (1 - r_{0} - r_{\rm g})^{k - 2} \cdot \frac {1}{2}
+ \sum _{s = 3}^{k} {k \choose s}
r_{\rm g}^{s} (1 - r_{0} - r_{\rm g})^{k - s} \cdot \frac {s}{2^{s - 1}} \right] \nonumber \\
& = \sum _{k = 2}^{+\infty} P(k)
{k \choose 2} r_{\rm g}^2 (1 - r_{0} - r_{\rm g})^{k - 2} \cdot \frac {1}{2}
+ \sum _{k = 3}^{+\infty} P(k) \sum _{s = 3}^{k}
{k \choose s} 
r_{\rm g}^{s} (1 - r_{0} - r_{\rm g})^{k - s} \cdot \frac {s}{2^{s - 1}} \nonumber \\
& = \frac {r_{\rm g}^2}{2}
\sum _{k = 2}^{+\infty} P(k) {k \choose 2} (1 - r_{0} - r_{\rm g})^{k - 2}
+ r_{\rm g} \sum _{k = 3}^{+\infty} P(k)
\sum _{s = 3}^{k} {k \choose s} s \left( \frac {r_{\rm g}}{2} \right)^{s - 1} (1 - r_{0} - r_{\rm g})^{k - s} \nonumber \\
& \equiv \frac {r_{\rm g}^2}{2}
\sum _{k = 2}^{+\infty} P(k) {k \choose 2} (1 - r_{0} - r_{\rm g})^{k - 2} + S_1.
\end{align}
In the last equation sign of above equations, we define the second summation as $S_1$.

From the definition of excess degree distribution $Q(k)$, we know that $kP(k) = cQ(k)$. Here we consider a more general form $P(k) {k \choose s}$ with $s \geqslant 1$.
We have
\begin{align}
\label{eq:Pk-k}
P(k) {k \choose s} s
& = P(k) \frac {k!}{s! (k - s)!} s \nonumber \\
& = P(k) \frac {k \cdot (k - 1)!}{(s  - 1)! (k - s)!} \nonumber \\
& = P(k) k \cdot \frac {(k - 1)!}{(s - 1)! (k - s)!} \nonumber \\
& = c Q(k)  {k - 1 \choose s - 1}.
\end{align}

We have
\begin{align}
S_1
& = r_{\rm g} \sum _{k = 3}^{+\infty}
\sum _{s = 3}^{k} c Q(k) {k - 1 \choose s - 1} \left( \frac {r_{\rm g}}{2} \right)^{s - 1} (1 - r_{0} - r_{\rm g})^{k - s} \nonumber \\
& = c r_{\rm g} \sum _{k = 3}^{+\infty} Q(k)
\sum _{s = 3}^{k} {k - 1 \choose s - 1} \left( \frac {r_{\rm g}}{2} \right)^{s - 1} (1 - r_{0} - r_{\rm g})^{k - s} \nonumber \\
& \equiv c r_{\rm g} \sum _{k = 3}^{+\infty} Q(k)
\sum _{s = 3}^{k} T_{2}^{(s)}.
\end{align}
In the last equation sign of above equations, we define the term in the second summation as $T_{2}^{(k)}$.
Then we have
\begin{align}
\label{eq:S2T3}
S_1
= c r_{\rm g} \sum _{k = 3}^{+\infty} Q(k)
\left[ \sum _{s = 1}^{k} T_{2}^{(s)} - T_{2}^{(1)} - T_{2}^{(2)} \right].
\end{align}
After some simple calculation, we have
\begin{align}
\sum _{s = 1}^{k} T_{2}^{(s)}
& = \sum _{s = 1}^{k} {k - 1 \choose s - 1}
\left( \frac {r_{\rm g}}{2} \right)^{s - 1} (1 - r_0 - r_{\rm g})^{k - s} 
= \left( 1 - r_0 - \frac{r_{\rm g}}{2} \right)^{k - 1}, \\
T_{2}^{(1)}
& = {k - 1 \choose 0}
\left( \frac {r_{\rm g}}{2} \right)^{0} (1 - r_0 - r_{\rm g})^{k - 1} 
= (1 - r_0 - r_{\rm g})^{k - 1}, \\
T_{2}^{(2)}
& = {k - 1 \choose 1}
\left( \frac {r_{\rm g}}{2} \right)^{1} (1 - r_0 - r_{\rm g})^{k - 2} 
= \frac {r_{\rm g}}{2} (k - 1) (1 - r_0 - r_{\rm g})^{k - 2}.
\end{align}
We have
\begin{align}
\sum _{s = 3}^{k} T_{2}^{(s)}
= \left( 1 - r_0 - \frac{r_{\rm g}}{2} \right)^{k - 1}
- (1 - r_0 - r_{\rm g})^{k - 1}
- \frac {r_{\rm g}}{2} (k - 1) (1 - r_0 - r_{\rm g})^{k - 2}.
\end{align}
Then, with equation (\ref{eq:S2T3}), we have
\begin{align}
S_1
& = c r_{\rm g} \sum _{k = 3}^{+\infty} Q(k)
\left[ 
 \left( 1 - r_0 - \frac{r_{\rm g}}{2} \right)^{k - 1}
- (1 - r_0 - r_{\rm g})^{k - 1}
- \frac {r_{\rm g}}{2} (k - 1) (1 - r_0 - r_{\rm g})^{k - 2} \right] \nonumber \\
& = c r_{\rm g} \sum _{k = 3}^{+\infty} Q(k)
\left[ \left( 1 - r_0 - \frac{r_{\rm g}}{2} \right)^{k - 1}
- (1 - r_0 - r_{\rm g})^{k - 1} \right] 
 - \frac {c r_{\rm g}^2}{2} \sum _{k = 3}^{+\infty} Q(k)
(k - 1) (1 - r_0 - r_{\rm g})^{k - 2} \nonumber \\
& \equiv cr_{\rm g} \sum _{k = 3}^{+\infty} Q(k) T_{3}^{(k)}
- \frac {c r_{\rm g}^2}{2} \sum _{k = 3}^{+\infty} Q(k) T_{4}^{(k)}.
\end{align}
In the last equation sign of above equations, we define two short-hand summations $T_{3}^{(k)}$ and $T_{4}^{(k)}$. Correspondingly, we have
\begin{align}
S_1
= cr_{\rm g} \left[ \sum _{k = 1}^{+\infty} Q(k) T_{3}^{(k)} - Q(1) T_{3}^{(1)} - Q(2) T_{3}^{(2)} \right]
- \frac {c r_{\rm g}^2}{2} \left[ \sum _{k = 2}^{+\infty} Q(k) T_{4}^{(k)} - Q(2) T_{4}^{(2)} \right].
\end{align}
We have
\begin{align}
T_{3}^{(1)}
& = 1 - 1 = 0, \\
T_{3}^{(2)}
& = \left( 1 - r_0 - \frac{r_{\rm g}}{2} \right)^{1} - (1 - r_0 - r_{\rm g})^{1} = \frac {r_{\rm g}}{2},\\
T_{4}^{(2)}
& = (2 - 1) \cdot 1 = 1.
\end{align}
We have
\begin{align}
c r_{\rm g} \left[ - Q(1) T_{3}^{(1)} - Q(2) T_{3}^{(2)} \right]
+ \frac {c r_{\rm g}^2}{2} Q(2) T_{4}^{(2)}
& = c r_{\rm g} \left[ - Q(1) \cdot 0 - Q(2) \cdot \frac {r_{\rm g}}{2} \right]
+ \frac {c r_{\rm g}^2}{2} Q(2) \cdot 1 \nonumber \\
& = 0.
\end{align}
Thus, we have
\begin{align}
S_1
& = cr_{\rm g} \sum _{k = 1}^{+\infty} Q(k) T_{3}^{(k)}
- \frac {c r_{\rm g}^2}{2} \sum _{k = 2}^{+\infty} Q(k) T_{4}^{(k)} \nonumber \\
& = cr_{\rm g}  \sum _{k = 1}^{+\infty} Q(k)
\left[ \left( 1 - r_0 - \frac {r_{\rm g}}{2} \right)^{k - 1}
- (1 - r_0 - r_{\rm g})^{k - 1} \right] 
- \frac {c r_{\rm g}^2}{2} \sum _{k = 2}^{+\infty} Q(k) (k - 1) (1 - r_0 - r_{\rm g})^{k - 2}.
\end{align}

For $P_5$, we finally have
\begin{align}
P_5
& = \frac {r_{\rm g}^2}{2} \sum _{k = 2}^{+\infty} P(k) {k \choose 2} (1 - r_0 - r_{\rm g})^{k - 2} \nonumber \\
& + cr_{\rm g}  \sum _{k = 1}^{+\infty} Q(k)
\left[ \left( 1 - r_0 - \frac {r_{\rm g}}{2} \right)^{k - 1}
- (1 - r_0 - r_{\rm g})^{k - 1} \right] 
- \frac {c r_{\rm g}^2}{2} \sum _{k = 2}^{+\infty} Q(k) (k - 1) (1 - r_0 - r_{\rm g})^{k - 2} \nonumber \\
& = \frac {r_{\rm g}^2}{2} \sum _{k = 2}^{+\infty}
\frac {c}{2} Q(k) (k - 1) (1 - r_0 - r_{\rm g})^{k - 2} \nonumber \\
& + cr_{\rm g}  \sum _{k = 1}^{+\infty} Q(k)
\left[ \left( 1 - r_0 - \frac {r_{\rm g}}{2} \right)^{k - 1}
- (1 - r_0 - r_{\rm g})^{k - 1} \right] 
- \frac {c r_{\rm g}^2}{2} \sum _{k = 2}^{+\infty} Q(k) (k - 1) (1 - r_0 - r_{\rm g})^{k - 2} \nonumber \\
& = cr_{\rm g}  \sum _{k = 1}^{+\infty} Q(k)
\left[ \left( 1 - r_0 - \frac {r_{\rm g}}{2} \right)^{k - 1}
- (1 - r_0 - r_{\rm g})^{k - 1} \right]
- \frac {c r_{\rm g}^2}{4} \sum _{k = 2}^{+\infty} Q(k) (k - 1) (1 - r_0 - r_{\rm g})^{k - 2}.
\end{align}
In the second equation sign of above equations, we adopt equation (\ref{eq:Pk-k}) with $s = 2$.
Thus we have equation (\ref{eq:P5-simplified}) in the main text.

For $R_{\ast}$, we finally have
\begin{align}
\label{eq:Ra-appendix}
R_{\ast}
& = P_3 +  P_4 + P_5 \nonumber \\
& = r_0 \sum _{k = 1}^{+\infty} P(k) k (1 - r_0 - r_{\rm g})^{k - 1} \nonumber \\
& + 2 r_0 \sum _{k = 1}^{+\infty} P(k) k
\left[ \left( 1 - r_0 - \frac {r_{\rm g}}{2} \right)^{k - 1} - (1 - r_0 - r_{\rm g})^{k - 1} \right] \nonumber \\
& + cr_{\rm g}  \sum _{k = 1}^{+\infty} Q(k)
\left[ \left( 1 - r_0 - \frac {r_{\rm g}}{2} \right)^{k - 1}
- (1 - r_0 - r_{\rm g})^{k - 1} \right]
- \frac {c r_{\rm g}^2}{4} \sum _{k = 2}^{+\infty} Q(k) (k - 1) (1 - r_0 - r_{\rm g})^{k - 2} \nonumber \\
& = 2 r_0 \sum _{k = 1}^{+\infty} P(k) k
\left( 1 - r_0 - \frac {r_{\rm g}}{2} \right)^{k - 1} 
- r_0 \sum _{k = 1}^{+\infty} P(k) k
(1 - r_0 - r_{\rm g})^{k - 1} \nonumber \\
& + cr_{\rm g}  \sum _{k = 1}^{+\infty} Q(k)
\left[ \left( 1 - r_0 - \frac {r_{\rm g}}{2} \right)^{k - 1}
- (1 - r_0 - r_{\rm g})^{k - 1} \right]
- \frac {c r_{\rm g}^2}{4} \sum _{k = 2}^{+\infty} Q(k) (k - 1) (1 - r_0 - r_{\rm g})^{k - 2} \nonumber \\
& = 2 r_0 \sum _{k = 1}^{+\infty} c Q(k)
\left( 1 - r_0 - \frac {r_{\rm g}}{2} \right)^{k - 1} 
- r_0 \sum _{k = 1}^{+\infty} c Q(k)
(1 - r_0 - r_{\rm g})^{k - 1} \nonumber \\
& + cr_{\rm g}  \sum _{k = 1}^{+\infty} Q(k)
\left[ \left( 1 - r_0 - \frac {r_{\rm g}}{2} \right)^{k - 1}
- (1 - r_0 - r_{\rm g})^{k - 1} \right]
 - \frac {c r_{\rm g}^2}{4} \sum _{k = 2}^{+\infty} Q(k) (k - 1) (1 - r_0 - r_{\rm g})^{k - 2} \nonumber \\
& = (2 c r_0 + c r_{\rm g}) \sum _{k = 1}^{+\infty} Q(k)
\left( 1 - r_0 - \frac {r_{\rm g}}{2} \right)^{k - 1} 
- (c r_0 + c r_{\rm g}) \sum _{k = 1}^{+\infty} Q(k)
(1 - r_0 - r_{\rm g})^{k - 1} \nonumber \\
& - \frac {c r_{\rm g}^2}{4} \sum _{k = 2}^{+\infty} Q(k) (k - 1) (1 - r_0 - r_{\rm g})^{k - 2}.
\end{align}
Thus we have equation (\ref{eq:Ra-simplified}) in the main text.

\end{widetext}

\section*{Appendix B: The theory based on the GLR procedure for the MVC problem}

\setcounter{equation}{0}
\renewcommand{\theequation}{B.\arabic{equation}}

This appendix is based on the main text of \cite{Zhao.Zhou-JSTAT-2019}.
We sketch here basic equations of energy density of the MVC problem based on the GLR procedure.

We consider a large sparse random graph $G = \{V, E\}$ with a vertex set $V$ and an edge set $E$.
We then adopt the cavity method to analytically calculate the fractions of vertices in its core $n$ and the fraction of roots $w$.
From the viewpoint of cavity method, $n$ and $w$ are marginal probabilities of a vertex to be in a core or be a root in $G$, respectively.
On a randomly chosen edge $(i, j) \in E$ between vertices $i$ and $j$,  from $i$ to $j$ we define two cavity probabilities: $\alpha$ as the probability that $j$ is a leaf, thus $i$ is the corresponding root; $\beta$ as the probability that $j$ is a root, while $i$ is not its corresponding leaf.
With the Bethe-Peierls approximation on sparse graphs \cite{Mezard.Montanari-2009}, the marginals $n$ and $w$ can be established with cavity probabilities as
\begin{align}
\label{eq:core-n}
n
& = \sum _{k = 2}^{+ \infty} P(k) \sum _{s = 2}^{k} {k \choose s} (1 - \alpha - \beta)^{s} \beta ^{k - s}, \\
\label{eq:core-w}
w
& = 1 - \sum _{k  = 0}^{+ \infty} P(k) (1 - \alpha)^{k} - \frac {1}{2} c \alpha ^2.
\end{align}
During the GLR procedure, a root and at least one leaf emerge at the same time.
Thus the cavity probabilities $\alpha$ and $\beta$ can be expressed as coupled equations as
\begin{align}
\label{eq:core-alpha}
\alpha
& = \sum _{k = 1}^{+ \infty} Q(k) \beta ^{k - 1}, \\
\label{eq:core-beta}
\beta
& = 1 - \sum _{k = 1}^{+ \infty} Q(k) (1 - \alpha)^{k - 1}.
\end{align}
For a graph ensemble or instance with a degree distribution $P(k)$, we first calculate the stable fixed $(\alpha, \beta)$ with equations (\ref{eq:core-alpha}) and (\ref{eq:core-beta}), and then calculate $n$ with equation (\ref{eq:core-n}) and $w$ with equation (\ref{eq:core-w}).

For equations (\ref{eq:core-alpha}) and (\ref{eq:core-beta}), when $c$ is small, we have a single and also stable fixed point $(\alpha, \beta)$ with $1 - \alpha - \beta = 0$. Correspondingly, there is no core after the GLR procedure.
The set of roots is simply a MVC configuration, thus $w$ is the energy density of the MVC problem.

When $c$ is large enough, there are three branches for the fixed point $(\alpha, \beta)$: $1 - \alpha - \beta > 0$ as the stable and also the physical solution, $1 - \alpha - \beta = 0$ as the trivial solution, and $1 - \alpha - \beta < 0$ as the unphysical solution.
To calculate the nontrivial $n$ and $w$ in the percolation theory, we choose the branch with $1 - \alpha - \beta > 0$.
This corresponds to the emergence of a core.
From the perspective of finding MVC configurations, the core needs further algorithm treatments.
Yet from the perspective of the theoretical approximation, we can still estimate the corresponding energy density for the MVC problem.
Here we simply assume that there is always no core in a graph.
We follow the trivial solution with $1 - \alpha - \beta = 0$, and the corresponding $w$ is the estimation of the energy density of the MVC problem.
By setting $1 - \alpha - \beta = 0$, we simplify equations (\ref{eq:core-alpha}) and (\ref{eq:core-beta}) into one self-consistent equation as
\begin{align}
\label{eq:core0-alpha}
\alpha
& = \sum _{k = 1}^{+ \infty} Q(k) (1 - \alpha) ^{k - 1}.
\end{align}
The right-hand side of the above equation is a monotonously decreasing function of $\alpha \in [0, 1]$. Thus there is only one fixed point of $\alpha$. Beware that when there is a core percolation, the only fixed solution is not stable.
By solving the fixed point of $\alpha$ with equation (\ref{eq:core0-alpha}), we calculate $w$ with equation (\ref{eq:core-w}) as the prediction of energy density $x$ of the MVC problem whether a graph has a core or not.


\end{document}